\documentclass[pra,aps,showpacs,superscriptaddress,twocolumn]{revtex4}
\usepackage{graphicx}
\usepackage{amsmath}
\usepackage{amsfonts}
\usepackage{amssymb}
\usepackage{epsfig}
\usepackage{color}

\begin{document}

\newcommand{\bra}[1]{\left\langle#1\right\vert}
\newcommand{\ket}[1]{\left\vert#1\right\rangle}

\title{Dicke model and environment-induced entanglement in ion--cavity QED}

\author{K. H\"ark\"onen}
\email{kari.harkonen@utu.fi}
\affiliation{Department of Physics and Astronomy, University of Turku, FI-20014 Turku, Finland}
\author{F. Plastina}
\affiliation{Dip. Fisica, Universit\`a della Calabria, and INFN -
Gruppo collegato di Cosenza, Arcavacata di Rende, Cosenza 87036, Italy}
\author{S. Maniscalco}
\affiliation{Department of Physics and Astronomy, University of Turku, FI-20014 Turku, Finland}

\date{\today}

\begin{abstract}
We investigate realistic experimental conditions under which
the collective Dicke model can be implemented in the ion-cavity
QED context. We show how ideal subradiance and superradiance can
be observed and we propose an experiment to generate entanglement
exploiting the existence of the subradiant state. We explore the
conditions to achieve optimal entanglement generation and we show
that they are reachable with current experimental technology.
\end{abstract}


\maketitle

\section{Introduction}

The Dicke model describes the dynamics of $N$ identical two-level
atoms interacting with a quantized three-dimensional
electromagnetic (EM) field \cite{Dicke}. Under certain conditions
the model predicts that the atoms interact with the quantized EM field collectively, giving rise to the widely studied phenomena of
superradiance and subradiance \cite{HarochePR,tanas02}. In free
space ideal superradiance and subradiance take place in the so
called small sample limit, i.e., when the atoms are so close to
each other that one can ignore any effect resulting from  their
different spatial positions. In this case the atoms are
indistinguishable with respect to their emission and absorption
properties; hence, the presence of equivalent paths through which
the emission process may occur gives rise to fully constructive
(superradiance) or destructive (subradiance) interference.

Ideal superradiance or subradiance in free space is very difficult to
observe in the experiments since it requires that the atoms are
placed in a regular pattern within a sample smaller than the
wavelength of the EM field they interact with (small sample case).
The requirement of a regular pattern is due to the presence of the
dipole-dipole forces that would otherwise break the symmetry under
permutation of any two atoms necessary to observe
superradiant-subradiant behavior. Such a regularity can be
achieved, e.g., with trapped-ion crystals \cite{IonsRev} or atoms
in optical lattices \cite{Bloch08}. In these systems, however, the
separation between the particles is typically larger or on the same
order of magnitude than the resonant wavelength (large sample
case). In the large sample case, cooperative effects still occur
but the subradiant state is not completely decoupled from the
dynamics. Indeed, partial subradiance and superradiance have been 
observed with trapped ions \cite{DeVoe}.

A way for relaxing the requirement for configuration regularity is
to place the small sample in a cavity resonator. In this case,
indeed, due to the Purcell effect, the cooperative atomic behavior
can be observed at much lower atomic density than in free space,
making the van der Waals dephasing caused by the irregular atomic
configuration negligible \cite{HarochePR}. Experiments observing
superradiance in the small sample case in a cavity have been
performed with Rydberg atoms \cite{HarocheSuperR}, giving results
in a very good agreement with the predictions of the single-mode
superradiance theory. In this experiment, all of the atoms are
equivalently coupled to the quantized mode of the EM field
(homogeneous case).

Recent advances in ion-cavity QED experiments make it possible to
confine arrays of ions inside an optical cavity in a regime in
which the width of their wave packet in position space is smaller
than the wavelength of the cavity mode they interact with
(Lamb--Dicke regime) \cite{Walther,Piety}. Moreover, it is
possible to accurately manipulate the position of the single ions
with respect to the intensity profile of the standing cavity mode, thereby allowing us to change the strength of the
coupling between each ion and the quantized EM field.

It has been  demonstrated theoretically that, when the atoms are
coupled with different strengths to the EM field, ideal
superradiance or subradiance can still occur, depending on the
particular spatial distribution of the atoms
\cite{Benivegna,BenivegnaPL,BuzekZei}. However, no experiments
have up to now confirmed these predictions by the inhomogeneous Dicke model. Very recently, an important step in this direction has been achieved at the University of Aarhus, where a collective strong coupling between an ion crystal and a cavity mode was observed \cite{Herskind2009}. In this paper, we
investigate in detail how the inhomogeneous single-mode Dicke
model (or Tavis-Cummings model \cite{TCmodel}) can be realized in
the ion-cavity QED context and the conditions under which
subradiance and superradiance can be observed.

Besides the importance in the study of fundamentals of quantum
theory, the realization of the Dicke model and the generation of
the subradiant state play a crucial role in quantum information
technology and quantum communication. Indeed, arrays of ions are
ideal candidates for quantum registers and their controlled
interaction with photons allows us to realize atom-light quantum
interfaces \cite{Kimble} and to distribute entanglement to
different nodes of quantum networks. The importance of the
subradiant states in this context stems from the fact that they
are robust entangled atomic states since they are completely
decoupled from the EM field.

The aim of this work is to discuss a realistic setup that is able
to show the collective behavior of trapped ions in a cavity. In
particular, since in the experiments performed so far the ions are
coupled to the EM mode via a Raman scheme in a $\Lambda$-configuration, we will include the entire level structure, which
is important in order to understand the decohering role of the spontaneous
emission from the upper and essentially unpopulated level. We
will also include cavity losses in order to study in detail the
deviation from the ideal cooperative Dicke model and to identify
the parameter regions in which such deviations are as small as
possible.

In fact, during the last two decades, several theoretical papers
have discussed issues such as entanglement generation, preparation
of nonclassical states, or realization of quantum gates in the
ion-cavity QED context assuming that the conditions to realize an
ideal Tavis-Cummings model were met
\cite{Pellizzari,vanEnk,Plenio,Zheng,Pachos,Lougovski,Chimczak,Li,Li07,Chimczak08,Bina}.
Thus, either the spontaneous emission or the cavity losses (or both
processes) are usually neglected \cite{vanEnk,Zheng,Li,Li07}.
Concerning spontaneous emission, for example, the assumption is
made that the emission rate is much smaller than the cavity
coupling constant
\cite{Pellizzari,Pachos,Lougovski,Chimczak,Chimczak08}. However,
this condition is not met in the ion-cavity QED experiments
\cite{Walther,Piety}. Furthermore, as we will demonstrate in this
paper, if one deals with simplified atomic level structures
\cite{Plenio,Zheng,Bina,Natali2007}, it is not possible to single out those
regions in parameter space for which the systems of trapped ions
behave collectively.

In this paper, we will take both the cavity losses and the spontaneous
emissions into account and employ $\Lambda$-type schemes to describe
the ions and to identify the experimental conditions under which
the coherent dynamics predicted by the single-mode Dicke model is
dominant with respect to losses and decoherence. This will also
allow us to present realistic protocols for entanglement
generation and to discuss ways to optimize the generated
entanglement using specific features of the trapped-ion system,
such as the ability to manipulate in a controlled way the relative
coupling between the ions and the cavity field.

The structure of the paper is the following. In
Sec.~\ref{sec:DickeModel} we review the properties of the
inhomogeneous single-mode Dicke model. In
Sec.~\ref{sec:effectiveModel} we present the Hamiltonian for two
ions in a cavity and we make the connection to the Dicke model by deriving an effective model describing the dynamics under realistic experimental conditions. Section~\ref{sec:resonantRegime} is devoted to the description of
the experimental proposal to observe subradiance and verify the
inhomogeneous Dicke model. Furthermore, in Sec.~\ref{sec:dispersiveRegime}
we explore another way to optimize the entanglement generation by
using off-resonant transitions. Finally, a summary of the results
and the conclusions are given in Sec.~\ref{sec:summary}.

\section{\label{sec:DickeModel}
Inhomogeneous single-mode Dicke model}

\subsection{Ideal cavity}

The single-mode Dicke model, or Tavis-Cummings model, is the
simplest quantum-mechanical model describing collective effects
such as superradiance and subradiance in cavity. It describes the
quasi-resonant interaction between $N$ identical two-level atoms
and a single quantized cavity mode. The Tavis-Cummings
Hamiltonian is
\begin{align}
H_\textrm{D} =&\,\, \omega_C \left( a^{\dag} a + \frac{1}{2} \right) + \sum_{j=1}^N  \omega_A \sigma_+^{(j)} \sigma_-^{(j)} \nonumber \\
& +\sum_{j=1}^N \left( \alpha^{(j)} a^{\dag} \sigma_-^{(j)} + \alpha^{(j)*} a \sigma_+^{(j)}\right),  \label{eq:HDicke2}
\end{align}
where $\omega_C$ and $\omega_A$ are the frequencies of the cavity
mode and the atomic transition, respectively; $a$ and
$a^{\dag}$ are the annihilation and the creation operators for the
cavity mode; and $ \sigma_-^{(j)} =\vert 0^{(j)} \rangle \langle
1^{(j)} \vert$ and $ \sigma_+^{(j)}=(\sigma_-^{(j)})^{\dag}$ are
the lowering and the raising operators for the $j$th atom, $\vert
0^{(j)} \rangle$ and $\vert 1^{(j)} \rangle$ being its ground and
excited states, respectively. Finally, $\alpha^{(j)}$ is the
coupling strength of the $j$th atom with the cavity field.
Inhomogeneity of the coupling strengths originates from
different relative positions of the atoms with respect to the
intensity profile of the standing EM mode supported by the cavity
resonator.

This model assumes that the cavity is ideal, as photon escape is
not taken into account, and that atomic spontaneous emission from the
excited to the ground state is negligible. The model also neglects
the atomic motion as well as recoil effects due to the absorption
and subsequent re-emission of a photon by the atoms. Moreover, the
dipolar coupling of the atoms and the EM field is expressed within
a rotating wave approximation (RWA), thereby suppressing the
non-energy-conserving terms. Finally, it implicitly assumes that
the coupling between the atoms and the cavity mode does not
change, i.e., that the atoms are kept at fixed positions. While
the RWA has been proven to work extremely well in optical
experiments, all other assumptions need further consideration. In
the following sections we will examine them in detail for the
ion-cavity QED setup.

\begin{figure}
\begin{center}
\includegraphics{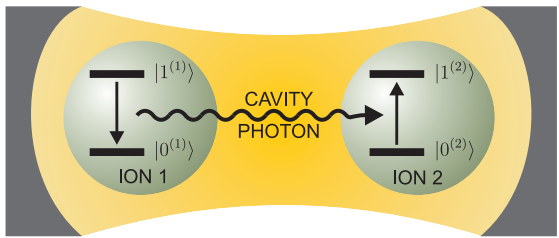}
\end{center}
\caption{\label{fig:dickeModel} (Color online) Two binary quantum
objects interacting through a quantized electromagnetic mode
supported by a cavity resonator. The dynamics of such an ideal system
is described by the Dicke model. }
\end{figure}

Using a suitable canonical transformation it has been shown that,
when only one excitation is present in the total system, the $N$
atoms interacting with the quantized field mode according to
Eq.~\eqref{eq:HDicke2} cooperate in such a way that only one
collective atomic mode (superradiant state) is coupled to the
field  \cite{BenivegnaPL}. Consequently, the energy exchange
between the atoms and the field can be completely suppressed if
the only field-coupled collective mode is unexcited.

For simplicity, we will from now on focus on the $N=2$ case
sketched in Fig.~\ref{fig:dickeModel}, and  we will denote the
energy eigenstates for the free ions as $|a^{(1)} b^{(2)}\rangle
\equiv |a^{(1)}\rangle\otimes |b^{(2)}\rangle$ (with $a,b=0,1$) and the corresponding Fock states of the cavity mode as $|n^{(C)}\rangle$, where $n= 0,1, \ldots$ .
The time evolution generated by $H_D$ is easily obtained
explicitly. For a cavity initially prepared in the vacuum state,
and in the presence of only one atomic excitation, the time
evolution of the amplitudes $c_{10} (t)$ and $c_{01} (t)$ to find the ions
in the states $\vert 1^{(1)} 0^{(2)} \rangle$ and $\vert 0^{(1)}
1^{(2)} \rangle$, respectively, is given by
\begin{align}
 c_{10}(t) =& \left[\, |r^{(2)}|^2 + |r^{(1)}|^2 \, {\cal E}(t)\,
\right]c_{10}(0) \nonumber \\
& - r^{(1)*} r^{(2)} \left[\, 1- {\cal E}(t)\, \right]c_{01}(0),
\label{eq:c1sing} \\
c_{01}(t) =& - r^{(1)} r^{(2)*} \left[\, 1- {\cal E}(t)\,
\right]c_{10}(0) \nonumber \\
& + \left[\, |r^{(1)}|^2 + |r^{(2)}|^2\, {\cal E}(t)\,
\right]c_{01}(0).
\label{eq:c2sing}
\end{align}
In the equations above the relative coupling strengths are defined as $r^{(j)}=\alpha^{(j)}/|\alpha_T |$, where $|\alpha_T| =
\sqrt{ |\alpha^{(1)}|^2 + |\alpha^{(2)}|^2 }$ is  the total
coupling strength, and
\begin{equation}
{\cal E}(t) = e^{ i \delta t /2} \Big[
\cos \Big( \frac{\Omega_\textrm{v} t}{2}\Big) - i\frac{
\delta}{\Omega_\textrm{v}} \, \sin \Big( \frac{\Omega_\textrm{v} t}{2}\Big)\Big],
\label{eq:Esing}
\end{equation}
where $\delta= \omega_A-\omega_C$ is the detuning and $\Omega_\textrm{v} = \sqrt{ 4  |\alpha_T|^2 + \delta^2}$ is
the \emph{vacuum Rabi frequency}. Note that $r^{(1)}$ and
$r^{(2)}$ are not independent parameters, since
$|r^{(2)}|=\sqrt{1-|r^{(1)}|^2}$.

The subradiant $ \ket{\psi_-}$ and the superradiant $ \ket{\psi_+}$ states are
\begin{align}
\big|\psi_- \big\rangle &= r^{(2)} \big| 1^{(1)} 0^{(2)} \big\rangle  - r^{(1)} \big| 0^{(1)} 1^{(2)} \big\rangle,
\label{eq:psim} \\
\big|\psi_+ \big\rangle &= r^{(1)*} \big| 1^{(1)} 0^{(2)} \big\rangle + r^{(2)*} \big|0^{(1)} 1^{(2)} \big\rangle,
\label{eq:psip}
\end{align}
and, in this case, they are position dependent through the
relative coupling strength parameters $r^{(1)}$ and $r^{(2)}$. As
one can see directly from Eq.~\eqref{eq:HDicke2}, the  state
$\ket{\psi_-} \otimes \ket{0^{(C)}}$ is an eigenstate of the
Tavis-Cummings Hamiltonian  with eigenvalue $\frac{1}{2}\omega_C + \omega_A$.
Therefore, when the atoms are prepared in this state, they are
completely decoupled from the cavity field and the system does not
evolve at all. In the case of equally strong couplings, i.e., for
$|r^{(1)}|=|r^{(2)}|=1/\sqrt{2}$, the subradiant and the superradiant
states coincide with the maximally entangled Bell states. In
general, however, these states are not maximally entangled.

\subsection{\label{sec:nonIdealCavity}
Non-ideal cavity}

We now proceed to generalize Eq.~\eqref{eq:HDicke2} to the case of
a lossy cavity. The imperfect reflectivity of the cavity mirrors
and consequent leakage of photons causes a Lorentzian broadening
of the spectral line corresponding to the mode supported by the ideal 
cavity. Accordingly, the microscopic atom-field interaction should
now take into account a continuum of modes described by a
Lorentzian distribution peaked at the central cavity frequency $\omega_C$.
For the sake of simplicity, and in view of the discussion in the
ion-cavity QED context, we restrict our attention to a one-dimensional cavity
model. Namely, we neglect the coupling with all the EM modes other
than the ones supported by the lossy cavity. In the rotating wave
approximation, the Hamiltonian is given by
\begin{align}
H =& \sum_{k} \omega_k \Big( a_k^{\dag}a_k + \frac{1}{2}\Big) + \sum_{j=1}^N \omega_A \sigma_+^{(j)}\sigma_-^{(j)} \nonumber \\
&+ \sum_{k} \sum_{j=1}^N \Big[ i g_k \sin \Big(\frac{\omega_k}{c} x^{(j)} \Big) a_k^\dagger \sigma_-^{(j)} + H.c. \Big], \label{eq:secondaeq}
\end{align}
where $a_k$ and $a^{\dag}_k$ are the annihilation and the creation
operators of cavity photons of frequency $\omega_k$, respectively.
Above, we have assumed that all the atoms have the same
electric dipole moment, which has been incorporated in the
coupling constants $g_k$, and we indicate with $x^{(j)}$ the
position of the atoms along the cavity axis. In the following we
will assume that each atom is kept at a fixed position inside the
cavity and that they are all well localized, i.e., the spread of
their wave function in position space is smaller than the
wavelength of the central cavity field mode: $\Delta x^{(j)} \ll
c/\omega_C$. Since all the significantly contributing modes are
close to the central mode (of frequency $\omega_C$), we have
\begin{equation}
\sin \left(\frac{\omega_k}{c} x^{(j)} \right) \simeq \sin \left(\frac{\omega_C}{c} x^{(j)} \right),
\end{equation}
and Eq.~\eqref{eq:secondaeq} takes the form
\begin{align}
H =& \sum_{k} \omega_k \Big( a_k^{\dag}a_k + \frac{1}{2}\Big) + \sum_{j=1}^N \omega_A \sigma_+^{(j)}\sigma_-^{(j)} \nonumber \\
&+ \sum_{j=1}^N \Big[ \chi^{(j)} \sigma_-^{(j)}  \sum_{k} g_k   a_k^\dagger + H.c. \Big], \label{eq:terzaeq}
\end{align}
with $\chi^{(j)} = i \sin (\omega_C x^{(j)}/c)$. In the continuum
limit the sum over the $k$-modes is replaced with an integral
$$
\sum_k |g_k|^2 \rightarrow \int \! d\omega J(\omega),
$$
where $J(\omega)$ is the reservoir spectral density. As mentioned
above, we assume a Lorentzian distribution for the spectrum of the
field inside the cavity; therefore, we take a spectral density of
the form
\begin{equation}
J(\omega) = \frac{W^2}{2 \pi} \frac{\kappa}{\left( \omega - \omega_C \right)^2 + (\kappa / 2)^2},
\label{eq:J}
\end{equation}
where the distribution is characterized by its full width at half
maximum value $\kappa$ and by a normalization parameter $W^2 =
\int \! d\omega \, J(\omega)$. Hence,  $\kappa$ describes the cavity
losses and $W$ describes the total coupling strength.

We focus again on the two-atom case, i.e., $N=2$, and we consider
the situation in which only one excitation is present in the total
atoms-field system. Starting from the Hamiltonian~\eqref{eq:secondaeq} and using the
Lorentzian spectral density~\eqref{eq:J}, it is possible to derive
an effective master equation
\begin{equation}
\frac{d \varrho}{dt} = - i \left[ H_D, \varrho \right]
-\frac{\kappa}{2} \left[ a^{\dag} a \varrho + \varrho a^{\dag} a -
2 a \varrho a^{\dag}\right] \label{eq:Dickeloss}
\end{equation}
for the dynamics of the atoms and the cavity mode of frequency
$\omega_C$ \cite{ESDLaura}. Here, $a$ and $a^{\dag}$ are the
annihilation and the creation operators for the central cavity mode,
which is damped at rate $\kappa$, and the coherent dynamics is
generated by $H_D$ in Eq.~\eqref{eq:HDicke2}, where the
coupling constants are identified as $\alpha^{(j)} = \chi^{(j)}
W$. From the exact solution of the effective master
equation~\eqref{eq:Dickeloss}, one can obtain the state of the
atomic system by tracing out the cavity degree of freedom: $\rho
(t) = \textrm{tr}_C [\varrho (t) ]$.

After performing the trace, and for an initially empty cavity, the
problem can be solved exactly. In the ordered basis $\left\{ \vert
1^{(1)} 1^{(2)} \rangle, \vert 1^{(1)} 0^{(2)} \rangle, \vert
0^{(1)} 1^{(2)} \rangle, \vert 0^{(1)} 0^{(2)} \rangle \right\}$,
the atomic density matrix can be written in the form \cite{Man08}
\begin{equation}
\rho(t) = 
\begin{pmatrix}
   0& 0 & 0 & 0 & \\ 0& |c_{10}|^2   &  c_{10} c_{01}^* & 0\\
 0& c_{10}^* c_{01} & |c_{01}|^2  & 0\\
   0&  0  & 0 & 1-|c_{10}|^2-|c_{01}|^2
\end{pmatrix}.
\label{eq:rhos}
\end{equation}
The dynamics of the two qubits is therefore completely characterized by the two amplitudes:
\begin{align}
c_{10}(t) =& \left[\, |r^{(2)}|^2 + |r^{(1)}|^2 \, {\cal E}(t)\,
\right]c_{10}(0) \nonumber \\
& -r^{(1)*} r^{(2)} \left[\, 1- {\cal E}(t)\, \right]c_{01}(0),
\label{eq:c1Sc1} \\
c_{01}(t) =& - r^{(1)} r^{(2)*} \left[\, 1- {\cal E}(t)\,
\right]c_{10}(0) \nonumber \\
& + \left[\, |r^{(1)}|^2 + |r^{(2)}|^2\, {\cal E}(t)\, \right]c_{01}(0),
\label{eq:c2Sc1}
\end{align}
with $r^{(j)} = \chi^{(j)} / |\chi_T |$, where $|\chi_T| = \sqrt{ |\chi^{(1)}|^2 + |\chi^{(2)}|^2 }$, and
\begin{eqnarray}
{\cal E}(t) = e^{-(\kappa -i 2\delta) t /4} \Big[
\cos \Big( \frac{\Omega_g t}{2}\Big) + \frac{\kappa- i 2
\delta}{2 \Omega_g} \sin \Big( \frac{\Omega_g t}{2}\Big) \Big], \nonumber \\
\label{eq:E}
\end{eqnarray}
where $\Omega_g = \sqrt{4 |\chi_T|^2 W^2 + \delta^2 +i
\delta \kappa -\kappa^2 / 4}$ is the \emph{generalized Rabi
frequency}. Note that Eqs.~\eqref{eq:c1Sc1} and \eqref{eq:c2Sc1} have
exactly the same structure as
Eqs.~\eqref{eq:c1sing} and \eqref{eq:c2sing}, obtained for the
single-mode Dicke model without losses. Formally, the cavity losses appear as an
additional imaginary part of the detuning $\delta \mapsto \delta + i\kappa
/2$. Accordingly, the effect of the cavity losses is described by
the modification of the time-dependent coefficient ${\cal E}(t)$,
which is now damped at rate $\kappa /4$, and by the
$\kappa$-dependent shift of the Rabi frequency. For $\kappa
\rightarrow 0$, the Lorentzian spectral density~\eqref{eq:J} tends
to Dirac's delta distribution, $J(\omega ) \rightarrow W^2 \delta
(\omega - \omega_C )$, and Eq.~\eqref{eq:E} reduces to
Eq.~\eqref{eq:Esing}, with $\alpha^{(j)} = \chi^{(j)} W$.

It is worth noticing that, as one sees directly from
Eq.~\eqref{eq:terzaeq}, the subradiant state $\vert \psi_-
\rangle$, given by Eq.~\eqref{eq:psim}, is still decoupled from
the vacuum cavity field. Hence, if the atomic system is initially
prepared in this state, no exchange of excitation with the cavity
field will take place.

\section{\label{sec:effectiveModel}
Effective model of ion-cavity interaction}

\subsection{Physical setup}

\begin{figure*}[tb]
\begin{center}
\includegraphics{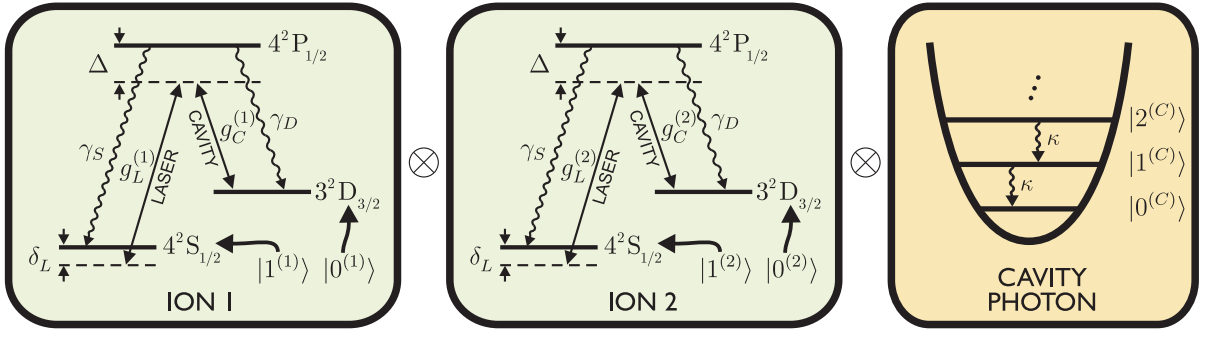}
\end{center}
\caption{\label{fig:threeLevelModel} (Color online) The relevant
electronic states of two identical $^{40}$Ca$^+$ ions and
corresponding couplings provided by an external pump laser and a
quantized cavity mode. The excited electronic state decays
spontaneously to the ground and to the metastable states, and the
cavity mode is damped as well.}
\end{figure*}

Ion-cavity QED experiments use calcium ions which are trapped in a linear
Paul microtrap and interact with a quantized mode of a
high-finesse optical cavity~\cite{Walther,Piety}. In
Fig.~\ref{fig:threeLevelModel} we show the relevant energy-level
structure, couplings, and decay channels for the compound system
of two $^{40}$Ca$^+$ ions and a single cavity mode. The atomic
ground state $4^2 S_{1/2}$ is coupled to the electronically
excited  state $4^2 P_{1/2}$ by a (classical) pump laser injected
from the side of the cavity. On the other hand, the excited state 
$4^2 P_{1/2}$ is coupled to a metastable state $3^2 D_{3/2}$ by the quantized cavity mode. The excited state
$4^2 P_{1/2}$ decays spontaneously to the states $4^2 S_{1/2}$
and $3^2 D_{3/2}$ at rates $\gamma_S$ and $\gamma_D$,
respectively, and the cavity photon is damped at rate $\kappa$,
as explained in the previous section.

A realistic theoretical description of the dynamics of a single
$^{40}$Ca$^+$ ion coupled to the cavity mode has been given in
Ref. \cite{Matthias}. The authors consider there also the effect
of cavity losses and spontaneous emission, taking into account all
the Zeeman sublevels of the three relevant  electronic states. The
main consequence of the presence of the Zeeman sublevels is a
reduction in the coupling driven by the cavity field by a factor of 
$\sqrt{3}$ with respect to the simpler three-level model
considered here. Therefore, we will use in the following a
three-level model scheme with such a reduced effective coupling to
account for the presence of the Zeeman sublevels. In the
experiments, the ions sit at the bottom of the trapping potential
and are cooled down to the Lamb-Dicke regime. Under these
conditions one can assume that the ions are kept at fixed
positions and neglect recoil during the emission-absorption
process.

In the following we will consider as initial atomic states those
in which one of the two atoms is in its ground state and the other
one is in its metastable state, i.e., the states $\ket{S^{(1)}
D^{(2)}}$ and $\ket{ D^{(1)} S^{(2)} }$. In order to prepare these
states, if the vibrational sidebands are not resolved, it is
necessary to use a selective laser  addressing of the individual
ions. This is routinely done in trapped-ion experiments with
$^{40}$Ca$^+$ ions (see, e.g., \cite{Blatt09}).

The two identical ions interact with the quantized cavity mode of
frequency $\omega_C$ via laser-assisted two-photon processes, as
shown in Fig.~\ref{fig:threeLevelModel}. The ions are irradiated
by a laser beam of frequency $\omega_L+\delta_L$. The laser beams
and the cavity field are far detuned by $\Delta$ from the
electronic level $\ket{P^{(j)}}$, such that $\omega_P - \omega_D =
\omega_C + \Delta$ and $\omega_P - \omega_S = \omega_L + \Delta$.
Therefore, the setup provides each ion $j=1,2$ with a Raman coupling between the levels $\ket{S^{(j)}}$ and $\ket{D^{(j)}}$.

The time evolution of the composite system of the two ions and the cavity mode can be described by a master equation
\begin{align}
\frac{d \varrho}{dt} =& -i \big[ H(t),\varrho \big] - \frac{\kappa}{2} \big(a^{\dag}a \varrho + \varrho a^{\dag}a-2a \varrho a^{\dag} \big) \nonumber \\
&- \frac{\gamma_S}{2} \sum_{j=1,2} \Big( A_{PP}^{(j)} \varrho + \varrho A_{PP}^{(j)} - 2 A_{SP}^{(j)} \varrho A_{PS}^{(j)} \Big) \nonumber \\
& - \frac{\gamma_D}{2} \sum_{j=1,2} \Big( A_{PP}^{(j)} \varrho + \varrho A_{PP}^{(j)} - 2 A_{DP}^{(j)} \varrho A_{PD}^{(j)} \Big),
\label{eq:model2me}
\end{align}
where we have included the cavity field damping at rate $\kappa$, the
spontaneous emission channels (two for each ion) at rates
$\gamma_S$ and $\gamma_D$, and where the coherent dynamics is
generated by a Hamiltonian
\begin{multline}
H(t) = \omega_C \Big( a^{\dag}a + \frac{1}{2} \Big) +  \sum_{j=1,2} \sum_{l=S,P,D}\omega_l A_{ll}^{(j)}   \\
+ \sum_{j=1,2}\Big( e^{-i (\omega_L+\delta_L)t} g^{(j)}_L A_{PS}^{(j)} + g_C^{(j)} a\, A_{PD}^{(j)} + H.c. \Big). \label{eq:model3H}
\end{multline}
The atomic operators are defined as $A_{ll'}^{(j)}=\vert l^{(j)}
\rangle \langle l'^{(j)} \vert$, with $l,l'=S,P,D$ and $j=1,2$.
Finally, the coherent couplings provided by the laser and the
cavity mode are, respectively,
\begin{align}
g_L^{(j)} &=\, \Omega\, e^{i k_L x^{(j)}}, \label{eq:gl} \\
g_C^{(j)} &=\, g \sin (k_C x^{(j)} ),  \label{eq:gc}
\end{align}
with $k_L$ and $k_C$ being the wave numbers of the laser and the standing cavity mode.

\subsection{\label{sec:effectiveProcesses}
Effective two-level model}

When the detuning $\Delta$ is sufficiently large compared to the
couplings, $\Delta \gg g_L^{(j)}, g_C^{(j)}$, the excited
electronic states $\ket{P^{(j)}}$ can be adiabatically eliminated
from the dynamics, as described in detail in Appendix~\ref{app:adiabaticElimination}. In this
case the system can be effectively described as composed of two
two-level atoms interacting with a cavity mode. For this purpose,
we denote the ground and the metastable states of the $j$th atom
as $\ket{1^{(j)}} \equiv \ket{S^{(j)}}$ and $\ket{0^{(j)}} \equiv
\ket{D^{(j)}}$ (N.B., the true atomic ground state corresponds to 
the excited state of the effective two-level system, since it is
able to emit a cavity photon through the Raman transition).

The adiabatic elimination of the excited levels $\{ |P^{(j)}\rangle \}$ is not at all
trivial due to the inclusion of the spontaneous emission
processes \cite{DiFidio}. We show in Appendix~\ref{app:adiabaticElimination} that an effective
Tavis-Cummings Hamiltonian can be derived, describing an
excitation exchange between the ions and the cavity. However, one
needs to include (i) two Stark shift terms per ion (one, in
particular, being dependent on the state of the cavity mode) and (ii) an overall
re-scaling  of both the free and the interaction energies by a
factor explicitly dependent on the emission rates.

It turns out that, in the interaction picture with respect to $H_0-\Delta \sum_{j}
A_{PP}^{(j)}$, where $H_0$ is given by the first two terms on the
right-hand side of Eq.~\eqref{eq:model3H}, the coherent part of the evolution of the ion-cavity system is described by an effective Hamiltonian
\begin{align}
H_{\rm eff} = & -\xi \sum_{j=1,2} \Big[ \Big( e^{-i \delta_L t} \frac{\beta^{(j)} g^* \Omega}{\Delta}\, a^{\dag} A_{01}^{(j)} + H.c. \Big) \nonumber \\
& +\frac{|\beta^{(j)}g|^2}{\Delta} \, a^{\dag} a A_{00}^{(j)} +
\frac{|\Omega|^2}{\Delta} A_{11}^{(j)} \Big], \label{eq:Heffion}
\end{align}
where the position-dependent parameters $\beta^{(j)}$ are defined
as
\begin{equation}
\beta^{(j)} = e^{i k_L x^{(j)}} \sin \big( k_C x^{(j)} \big),
\label{eq:coefficient}
\end{equation}
and the dimensionless renormalizing prefactor is
\begin{equation}
\xi =  \frac{\Delta^2}{\Delta^2 + (\gamma_S+\gamma_D)^2/4}.
\label{eq:xi}
\end{equation}
This Hamiltonian resembles the Tavis-Cummings Hamiltonian~\eqref{eq:HDicke2}, except for the photon-dependent Stark shift term. However, since the original microscopic model includes
dissipative processes, the unitary evolution generated by $H_{\rm
eff}$ needs to be supplemented by decohering terms that have a very peculiar structure. Indeed, the effective master equation that describes the time evolution of the ions and the cavity contains four (now both dissipative and non-dissipative) processes (described by jump operators) that take into account the effects of the spontaneous emission as seen in the
restricted atomic subspaces spanned by $\{|0^{(j)}\rangle, |1^{(j)}\rangle \}$. The cavity damping appears in the restricted subspace in the same form as in the original model. 

The effective master equation reads
\begin{align}
\frac{d \varrho}{dt} =& -i \left[ H_{\rm eff}, \varrho\right] - \frac{\kappa}{2} \left( a^{\dag}a \varrho + \varrho a^{\dag}a -2a \varrho a^{\dag} \right) \nonumber \\
& - \sum_{\substack{j=1,2\\m=S,D}} \frac{\Gamma_m^{(j)}}{2} \Big[ C_m^{(j)\dag}C_m^{(j)} \varrho +\varrho \, C_m^{(j)\dag}C_m^{(j)} \nonumber \\
& \qquad\qquad\qquad\, -2  C_m^{(j)} \varrho \, C_m^{(j)\dag} \Big],
\label{eq:meeffion}
\end{align}
where the jump operators for each ion $j$ are
given by
\begin{align}
C_S^{(j)} &= e^{-i \delta_L t} \Omega \, A_{11}^{(j)} + \beta^{(j)*} g\, a A_{10}^{(j)}, \label{eq:jumpC1} \\
C_D^{(j)} &= e^{-i \delta_L t} \Omega \, A_{01}^{(j)} + \beta^{(j)*} g\, a A_{00}^{(j)}, \label{eq:jumpC2}
\end{align}
while the effective decay rates are $\Gamma_m^{(j)} =\xi \gamma_m / \Delta^2$, where $m=S,D$ and the prefactor $\xi$ is given by Eq.~\eqref{eq:xi}. The structure of these jump operators is easy to interpret once
the full level configurations of Fig.~\ref{fig:threeLevelModel}
are taken into account. Let us consider, for example, the operator
$C_S^{(j)}$ of Eq.~(\ref{eq:jumpC1}). It arises from the spontaneous emission process $4^2 P_{1/2}\rightarrow 4^2 S_{1/2}$ of the $j$th atom, now being restricted to the two-dimensional subspace $\{|0^{(j)}\rangle,|1^{(j)}\rangle\}$. The jump operator $C_S^{(j)}$ has two contributions, both of
them describing non-dissipative decoherence by pure dephasing processes (as one understands from the fact that they do not produce any excitation loss). These two
contributions account for the interruption of the ion-cavity
excitation exchange (vacuum Rabi cycle) by the spontaneous
emission. The first term is an unwanted repopulation of state  $|1^{(j)}\rangle$ occurring after the laser has virtually brought the
system to the intermediate level $|P^{(j)}\rangle$ of the full Raman cycle. The
second term is also due to decay into state $|1^{(j)}\rangle$, but this time the virtual excitation of level $|P^{(j)}\rangle$ is caused by the cavity field. In conclusion, both processes interrupt the
vacuum Rabi cycle without the excitation being lost as, at the
end, the two-level system is found in its excited state $|1^{(j)}\rangle$. This
implies that the excitation exchange can restart, but with a different
phase. Thus, $C_S^{(j)}$ describes a phase error.

A similar interpretation scheme can be adopted for the two terms
constituting $C_D^{(j)}$ in Eq.~\eqref{eq:jumpC2}. However, this time the involved process is the spontaneous emission $4^2 P_{1/2}\rightarrow 3^2 D_{3/2}$. Whether it occurs after the
virtual excitation of level $|P^{(j)}\rangle$ performed by the laser (first term) or by the
cavity field (second term), the result is that at the end the two-level system is found in its ground state $|0^{(j)}\rangle$ and that one excitation has been lost either from the atom or from the cavity mode. Therefore, this jump operator causes dissipative decoherence. We note that, at this stage, the four jump operators of
Eqs.~\eqref{eq:jumpC1}-\eqref{eq:jumpC2} are both explicitly time
dependent and implicitly position dependent via the coefficients
$\beta^{(j)}$ [see Eq.~\eqref{eq:coefficient}].

A phase rotation within the restricted Hilbert space, spanned by
the states with at maximum one excitation, allows transforming
the effective Hamiltonian~\eqref{eq:Heffion} into the Tavis-Cummings
Hamiltonian~\eqref{eq:HDicke2} as well as removing simultaneously the
time dependence from the jump operators~\eqref{eq:jumpC1} and \eqref{eq:jumpC2}. This is described in Appendix~\ref{app:phaseRotation}. Therefore,
in a suitable rotating frame,  the following effective
Tavis-Cummings Hamiltonian is obtained:
\begin{align}
H_D^\textrm{eff} =&\,\, \omega_C^\textrm{eff} \Big( a^{\dag} a + \frac{1}{2} \Big) + \sum_{j=1,2}  \omega_A^\textrm{eff} \, \sigma_+^{(j)} \sigma_-^{(j)} \nonumber \\
& +\sum_{j=1,2} \Big( \alpha_\textrm{eff}^{(j)} \, a^{\dag} \, \sigma_-^{(j)} + \alpha_\textrm{eff}^{(j)*} \, a \, \sigma_+^{(j)}\Big), \label{eq:Hdeff}
\end{align}
where we have introduced again the spin inversion operators used in Sec.~\ref{sec:DickeModel}. The effective Dicke model parameters are
\begin{align}
\omega_C^\textrm{eff} & = -\xi \, \frac{2 |\beta_T g|^2}{3 \Delta}, \label{eq:omegaCEff}\\
\omega_A^\textrm{eff} & = \delta_L - \xi \Big( \frac{|\Omega|^2}{\Delta} - \frac{|\beta_T g|^2}{3\Delta} \Big) , \label{eq:omega0Eff}\\
\alpha_\textrm{eff}^{(j)} & = -\xi \, \frac{ \beta^{(j)} g^* \Omega }{\Delta} \equiv \beta^{(j)} g_\textrm{eff}, \label{eq:alphaEff}
\end{align}
where $|\beta_T | = \sqrt{|\beta^{(1)} |^2 + |\beta^{(2)} |^2 }$. The effective detuning is given by
\begin{equation}
\delta_\textrm{eff} = \omega_A^\textrm{eff} - \omega_C^\textrm{eff} = \delta_L -\xi \, \frac{ |\Omega|^2 - |\beta_T g|^2 }{\Delta }, \label{eq:deltaEff}
\end{equation}
and the relative effective coupling strengths $r^{(j)}$ [cf.~Eqs.~\eqref{eq:c1sing} and\eqref{eq:c2sing}] are directly given by the position-dependent parameters $\beta^{(j)}$,
since now $r^{(j)} = \alpha_\textrm{eff}^{(j)} /
|\alpha_{\textrm{eff},T}| = \beta^{(j)} / |\beta_T|$.

Comparing Eqs.~\eqref{eq:Hdeff} and \eqref{eq:meeffion} with
Eqs.~\eqref{eq:HDicke2} and \eqref{eq:Dickeloss}, respectively, we
see that, when the effective atomic spontaneous emissions are
negligible, this system allows us to realize the Dicke model in the
non-ideal cavity case.

\subsection{\label{sec:scaling}
Effective spontaneous emission processes}

As mentioned before, we restrict our study to the case in which
only one or zero quanta are present in the composite system of the two ions and the cavity mode.
Therefore, the compound state of the two atoms and the cavity
photon can be expressed in the basis  $\{ \ket{ 0^{(1)} 0^{(2)}
0^{(C)} }$, $\ket{ 0^{(1)} 0^{(2)} 1^{(C)} }$, $\ket{ 0^{(1)}
1^{(2)} 0^{(C)} }$, $\ket{ 1^{(1)} 0^{(2)} 0^{(C)} } \}$ (see
Appendix~\ref{app:phaseRotation}). Consequently, the jump operators~\eqref{eq:jumpC1}-\eqref{eq:jumpC2} can  be normalized with respect to the operator
norm $\|A\| = \sup_{\|\phi\|=1} \|A\ket{ \phi } \|$, where $\ket{
\phi }$ belongs to the Hilbert space spanned by the basis defined
above. The introduction of the normalized jump operators allows to
define the effective spontaneous emission decay rates
$\Gamma_m^{(j)}$ unambiguously.

The normalized jump operators are
\begin{align}
C_S^{(1)} &= |1^{(1)} 0^{(2)} 0^{(C)}\rangle \langle \Phi_1 |, \label{eq:scaledC11}\\
C_S^{(2)} &= |0^{(1)} 1^{(2)} 0^{(C)}\rangle \langle \Phi_2 |, \label{eq:scaledC12}\\
C_D^{(1)} &= |0^{(1)} 0^{(2)} 0^{(C)}\rangle \langle \Phi_1 |,\\
C_D^{(2)} &= |0^{(1)} 0^{(2)} 0^{(C)}\rangle \langle \Phi_2 |, \label{eq:scaledD12}
\end{align}
where the decaying states are
\begin{align}
|\Phi_1 \rangle & = \frac{ \Omega^* \, |1^{(1)} 0^{(2)} 0^{(C)}\rangle  + \beta^{(1)} g^*\, |0^{(1)} 0^{(2)} 1^{(C)}\rangle }{\sqrt{|\Omega |^2 + |\beta^{(1)} g|^2 } }, \label{eq:decayingState1} \\
|\Phi_2 \rangle & = \frac{ \Omega^* \, |0^{(1)} 1^{(2)} 0^{(C)}\rangle + \beta^{(2)} g^*\, |0^{(1)} 0^{(2)} 1^{(C)}\rangle }{\sqrt{|\Omega |^2 + |\beta^{(2)} g|^2 } }. \label{eq:decayingState2}
\end{align}
The corresponding rescaled decay rates are given by
\begin{align}
\Gamma_S^{(j)} &= \xi \, \big( |\Omega|^2 + |\beta^{(j)} g|^2 \big) \frac{ \gamma_S }{ \Delta^2 }, \label{eq:scaledDecayRateS}\\
\Gamma_D^{(j)} &= \xi \, \big( |\Omega|^2 + |\beta^{(j)} g|^2 \big) \frac{ \gamma_D }{ \Delta^2 }. \label{eq:scaledDecayRateD}
\end{align}
The cavity photon annihilation operator $a = |0^{(1)} 0^{(2)} 0^{(C)}\rangle \langle 0^{(1)} 0^{(2)} 1^{(C)} |$ is already normalized in our restricted basis.

The spontaneous emission decay rates for the considered states of
a calcium atom are $\gamma_S / 2 \pi = 22.3$ MHz and
$\gamma_D / 2 \pi = 1.7$ MHz. Therefore, $\Gamma_S^{(j)} \gg
\Gamma_D^{(j)}$ and the dominant effective spontaneous emission
jump processes are described by the operators
$C_S^{(j)}$. Consequently, according to the discussion above, the
main decoherence sources are the non-dissipative dephasing processes that conserve the energy of the ion--cavity system.

The character of the decaying state, and hence the corresponding
jump operator is defined by the balance between the strengths of the laser pumping $\Omega$ and the cavity coupling $\beta^{(j)} g$. In the \emph{strong laser pumping} case ($|\Omega |
\gg |\beta^{(j)} g|$) the non-unitary dynamics of the atomic
reduced system is dominated by phase diffusion processes described
by the operators $A_{11}^{(j)}$. In the \emph{weak laser pumping}
case ($|\Omega | \ll |\beta^{(j)} g|$), on the contrary, the
processes described by the operators $a A_{10}^{(j)}$
dominate. Moreover, as one can see from
Eqs.~\eqref{eq:decayingState1} and \eqref{eq:decayingState2}, one can
further modify the character of the specific atomic jump operators
by changing the relative position of the ions with respect to the
cavity field through the $\beta^{(j)}$ parameters.

The significance of the spontaneous emissions can be estimated by the ratio
\begin{equation}
\bigg| \frac{ \Gamma_S^{(j)} }{ \alpha_\textrm{eff}^{(j)} } \bigg| = \frac{ 1 + |\beta^{(j)} g / \Omega |^2 }{ | \beta^{(j)} g / \Omega | } \, \frac{ \gamma_S }{ \Delta }.
\label{eq:decayVersusCoupling}
\end{equation}
For a fixed detuning $\Delta$ this ratio has its minimum value $2
\gamma_S / \Delta $ when $|\beta^{(j)} g /  \Omega| = 1$, i.e.,
when the couplings provided by the laser and the cavity field are
equally strong. On the other hand, for fixed coupling strengths,
the ratio is inversely proportional to the detuning $\Delta$. This
can be exploited in order to minimize the role of the effective spontaneous decay. The
cavity damping $\kappa$ is neither affected by the detuning nor the couplings.

Finally, we note that for large detunings, $\Delta \gg \gamma_S, \gamma_D $, the dimensionless prefactor $\xi \sim 1$ and the effective decay rates as well as the effective coupling terms have simplified expressions. The effective couplings are then given by $\alpha_\textrm{eff}^{(j)} \sim -\beta^{(j)} g^* \Omega / \Delta$, while in the limit of strong and weak laser pumpings the dominating decay rates are $\Gamma_S^{(j)} \sim |\Omega|^2 \gamma_S / \Delta^2$ and $\Gamma_S^{(j)} \sim |\beta^{(j)} g|^2 \gamma_S / \Delta^2$, respectively.

\section{\label{sec:resonantRegime}
Environment-induced entanglement: Resonant regime}

In this section we study, analytically and numerically, the
dynamics of the entanglement between the electronic degrees of
freedom of the two atoms. The generation of entanglement between
the ions and its persistence at long times are, indeed, a clear
manifestation of the collective (subradiant) behavior. In
particular, entanglement generation is mediated by the interaction
with the quantized cavity field which is initially prepared in the
vacuum state. If the atomic spontaneous emission processes are negligible
and we face the bare Dicke model, the dynamics can be described
exactly. We compare these exact analytical results to numerical
simulations including the spontaneous emission effects. The
simulations were implemented by using the Monte Carlo wave
function (MCWF) method \cite{MCWF,ZollerCarmichael}. We begin by
considering the resonant case, where the effective detuning
$\delta_\textrm{eff} = 0$, with $\delta_\textrm{eff} $ given by
Eq.~\eqref{eq:deltaEff}.

\subsection{\label{sec:analytical}
Analytical solution neglecting spontaneous emission}

The effective model describing the dynamics when spontaneous emissions are negligible is given by the master equation~\eqref{eq:Dickeloss} with the effective Tavis--Cummings Hamiltonian~\eqref{eq:Hdeff}, as described in Sec.~\ref{sec:nonIdealCavity}. The analytical solution for the atomic density matrix is given by Eqs.~\eqref{eq:rhos} and \eqref{eq:E}, with $\chi^{(j)} W = \alpha_\textrm{eff}^{(j)} = \beta^{(j)} g_\textrm{eff}$.

We are interested in the collective dynamics when initially one excitation is present in the atomic system and the cavity is in its vacuum state. Any initial atomic state containing one excitation can be written in terms of the superradiant and subradiant states~\eqref{eq:psim}-\eqref{eq:psip} as
\begin{equation}
\ket{\psi(0) }= \beta_+  \ket{\psi_+}  + \beta_- \ket{\psi_-}. \label{eq:nonloso}
\end{equation}
As time passes, the collective atomic state decays via the evolution of the superradiant component,
\begin{equation}
\langle \psi_+ \vert \psi (t) \rangle = {\cal E}(t) \, \beta_+,
\end{equation}
with ${\cal E}(t)$ given by Eq.~\eqref{eq:E}. The subradiant component $\langle \psi_- \vert \psi (t) \rangle = \beta_-$, however, remains unchanged. Consequently, for times, such that $\kappa t \gg 1$, the atomic state will be in general a statistical mixture of the collective ground state $| 0^{(1)} 0^{(2)} \rangle$ and the subradiant state $\ket{\psi_-}$ with weights dependent on $\beta_-$, which in turn depends on the relative coupling strengths $r^{(j)}$.

In the following we focus on the dynamics of entanglement between the atoms. In order to quantify the stationary asymptotic entanglement of the final state we use Wootters's concurrence \cite{wootte} which, for a density matrix of the form of Eq.~\eqref{eq:rhos}, is given by
\begin{equation}
C(t) = 2 \left| c_{10}(t) c_{01}^*(t)\right|, \label{eq:concurrdef}
\end{equation}
with $c_{10}(t)$ and $c_{01}(t)$ given by Eqs.~\eqref{eq:c1Sc1} and \eqref{eq:c2Sc1}. In general, the concurrence is zero for factorized states and unity for maximally entangled states. For $\kappa t \gg 1$ we obtain a stationary concurrence value
\begin{equation}
C_\textrm{stat}=2 |r^{(1)}r^{(2)}| \left| \beta_- \right|^2.
\end{equation}
As expected, the value of the stationary concurrence is directly
related to the subradiant component of the initial state. If both
atoms are coupled to the EM field, the stationary value of the
concurrence, for any initial state with $\beta_- \ne 0$, will be
nonzero. When the atoms are initially prepared in the
superradiant state, i.e., $\beta_- =0$, the system approaches
asymptotically the pure factorized state $| 0^{(1)} 0^{(2)}
\rangle$.

For the initially factorized states $\ket{1^{(1)} 0^{(2)}}$ and
$\ket{0^{(1)} 1^{(2)}}$, the interaction with the environment
generates entanglement in the atomic system. For these initial
states the stationary concurrence takes the values $C_\textrm{stat} =
2 |r^{(1)}|(1- |r^{(1)}|^2)^{3/2} $ and $C_\textrm{stat} = 2
|r^{(1)}|^3 \sqrt{1- |r^{(1)}|^2}$, respectively. As we have
noticed in Ref. \cite{Man08}, the factorized states are those that
maximize the stationary concurrence for certain values of
$r^{(1)}$. The maximum value of stationary concurrence, for both
the two factorized initial states considered here, is
$C_\textrm{stat}^\textrm{max} = \max_{\,|r^{(1)}| \in [0,1]}
C_\textrm{stat} \simeq 0.65$. This value is obtained with
$|r^{(1)}|= 0.5$ and $|r^{(1)}| \simeq 0.87$ (i.e., $|r^{(2)}|=
0.5$) for initial states $\ket{1^{(1)} 0^{(2)}}$ and $\ket{0^{(1)}
1^{(2)}}$, respectively.

We note in passing that when only one of the two atoms is coupled
to the EM field, i.e., $r^{(1)}=0$ or $r^{(2)}=0$, the stationary
concurrence is zero. In this case, indeed, the subradiant and the
superradiant states coincide with states $\ket{1^{(1)} 0^{(2)}}$
and $\ket{0^{(1)} 1^{(2)}}$ as one can see from
definitions~\eqref{eq:psim} and \eqref{eq:psip}.

From the definition of the generalized Rabi frequency given by
Eq.~\eqref{eq:E}, which in the resonant case reads as
$\Omega_g = \sqrt{4 |\beta_T g_\textrm{eff}|^2 - \kappa^2
/ 4}$, two extreme regimes can be defined. In the \textsl{weak ion-cavity
coupling regime}, defined by  $4 |\beta_T g_\textrm{eff} |
\ll \kappa$,  the generalized Rabi frequency is purely imaginary.
Therefore, according to Eq.~\eqref{eq:E}, the Dicke model predicts
a solution given by monotonic hyperbolic sine and cosine
functions. The opposite limit is the \textsl{strong ion-cavity coupling regime}, defined by $4 |\beta_T g_\textrm{eff} | \gg
\kappa $. In this case the generalized Rabi frequency is real and
the Dicke model predicts damped oscillatory dynamics.

\subsection{MCWF simulations in the presence of spontaneous emission}

In this section, we focus on the effect of the spontaneous emissions on
the subradiant-state-based entanglement generation described in
the previous section. We consider again as initial atomic
state $|\psi (0) \rangle = |1^{(1)} 0^{(2)}\rangle$ with the cavity
in the vacuum state $|0^{(C)}\rangle$. For a given value of $r^{(1)} \in [0,1]$, we choose $\beta^{(1)}$ and $\beta^{(2)}$ to be positive real numbers such that the larger of the two is always unity and the smaller one is $\min \{ r^{(1)}/ \sqrt{ 1 - r^{(1)2}}, \sqrt{1 - r^{(1)2}}/ r^{(1)} \}$ [cf.~definition~\eqref{eq:coefficient}]. Now $|\beta_T|^2 = |\beta^{(1)}|^2 + |\beta^{(2)}|^2 = \textrm{min} \{ 1/r^{(1)2},1/ ( 1-r^{(1)2}) \} \in [1,2]$. The physical parameters
have been chosen in accordance to the experiments of Ref.
\cite{Walther} and  are summarized in
Table~\ref{tab:physicalValues}.  The size of the ensemble in the
MCWF simulations is $N=1000$. We are using the variant of MCWF
method described in \cite{ZollerCarmichael}.

\begin{table}[tb]
\caption{ \label{tab:physicalValues} Values of physical quantities
used in the simulations. Note that the cavity coupling is here
explicitly scaled by the Clebsch--Gordan factor $1/\sqrt{3}$ and,
in the text, also by the position-dependent parameters
$\beta^{(j)}$.}
\begin{center}
\begin{tabular}{l@{\quad}c@{\quad}r@{}l}
\hline
\hline
Quantity & Symbol & \multicolumn{2}{c}{Value (2$\pi$ MHz)} \\ \hline
Laser coupling & $\Omega$ & \qquad\quad 9&.0 \\
Cavity coupling & $g$ & 6&.5 $/ \sqrt{3}$\\
Decay rate $4^2 P_{1/2}\rightarrow 4^2 S_{1/2}$ & $\gamma_S$ & 22&.3 \\
Decay rate $4^2 P_{1/2}\rightarrow 3^2 D_{3/2}$ & $\gamma_D$ & 1&.7 \\
Detuning & $\Delta_0$ & 20&.0 \\
Cavity damping & $\kappa_0$ & 1&.2 \\
\hline
\hline
\end{tabular}
\end{center}
\end{table}

The value of the cavity coupling constant $g$ in
Table~\ref{tab:physicalValues} refers to  the new miniature trap
recently realized at the University of Sussex \cite{W}. The
reference value $\kappa_0$ for the cavity damping can nowadays be
improved by at least one order of magnitude. Finally, the detuning
$\Delta$ can be easily increased in the experiments, with respect
to the reference value $\Delta_0$.

With the experimental parameters of Table~\ref{tab:physicalValues}, the coupling strengths $\Omega$ and $g$ are of the same order. Therefore, neither the strong nor the weak laser pumping regimes, introduced
in Sec.~\ref{sec:scaling}, are reached and, consequently, all the effective
decay processes caused by the spontaneous emission are
combinations of two different physical operations, as interpreted in Sec.~\ref{sec:effectiveProcesses}. 

Let us denote the atomic density-matrix components as
$\rho_{ab,cd} \equiv \langle a^{(1)} b^{(2)} | \rho | c^{(1)}
d^{(2)} \rangle$, where $a,b,c,d=0,1$. The density matrix remains
still in the same block form of Eq.~\eqref{eq:rhos} even in the
presence of spontaneous emissions. The concurrence is therefore
given by $C(t) = 2 |\rho_{01,10}(t)|$.

\begin{figure}[tb]
\begin{center}
\includegraphics{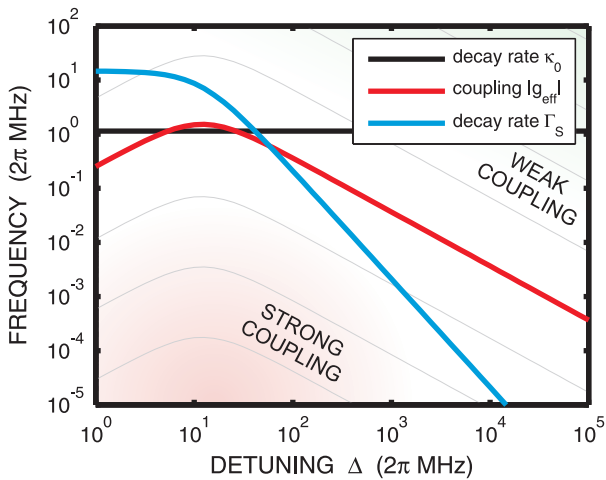}
\end{center}
\caption{\label{fig:scaling} (Color online) Scaling of the effective ion-cavity coupling $|g_\textrm{eff}|$ (middle line for large $\Delta$) and effective spontaneous emission decay rates $\Gamma_S^{(j)}$ (lowest line for large $\Delta$; with $\beta^{(j)}=1$) as a function of the detuning $\Delta$. The isocurves  $4 |\beta_T g_\textrm{eff}|/\kappa =$ const (thin lines) are parallel to the $|g_\textrm{eff}|$ curve, so that the weak ion-cavity coupling regime is in the upper right corner and the strong ion-cavity coupling regime in the lower left one. The cavity decay rate $\kappa$ (horizontal line) does not depend on the detuning. The effective spontaneous emission events are suppressed for large detunings.}
\end{figure}

In the following we will examine the effect of the spontaneous
emissions by comparing the concurrence as a function of time for
fixed values of $4 |\beta_T g_\textrm{eff}|/\kappa = 4 |\beta_T
\xi g \Omega  / \kappa \Delta|$. We study large detunings ($\Delta \gg \gamma_S, \gamma_D$), so the prefactor $\xi \sim 1$. In the examples we change $\kappa$ and $\Delta$, such that $\kappa /\kappa_0 = 0.1, 0.01$ and $\Delta / \Delta_0 = 10, 100, 1000$, while keeping the product $\kappa
\Delta$ constant. Physically, this corresponds to using different cavity qualities and detunings which, furthermore, influences the effective dynamical parameters. Larger
detunings, indeed, suppress the effective spontaneous emissions in
favor of the coherent dynamics, as explained in
Sec.~\ref{sec:scaling}. The situation is clarified in
Fig.~\ref{fig:scaling} which shows the scaling of the effective
coupling strength $g_\textrm{eff}$ and the dominant spontaneous
emission decay rate $\Gamma_S^{(j)}$ [cf.
Eqs.~\eqref{eq:alphaEff}, \eqref{eq:scaledDecayRateS}, and \eqref{eq:scaledDecayRateD}] as functions of detuning $\Delta$. The cavity damping rate $\kappa$ is not
affected by the detuning. The relative values of the three key
parameters $g_\textrm{eff}$, $\Gamma_S^{(j)}$, and $\kappa$
characterize the dynamical regime: (i) the ratio
$|g_\textrm{eff}|/\kappa$ defines the strong and the weak
ion-cavity coupling regimes and (ii) the magnitude of $\Gamma_S^{(j)}$ compared to
$|g_\textrm{eff}|$ and $\kappa$, in turn, describes the significance of the spontaneous emission processes and tells us whether the dynamics is well described by the Dicke model or not.

\subsubsection{\label{sec:subsub}Weak ion-cavity coupling regime}

\begin{figure}[tb]
\begin{center}
\includegraphics{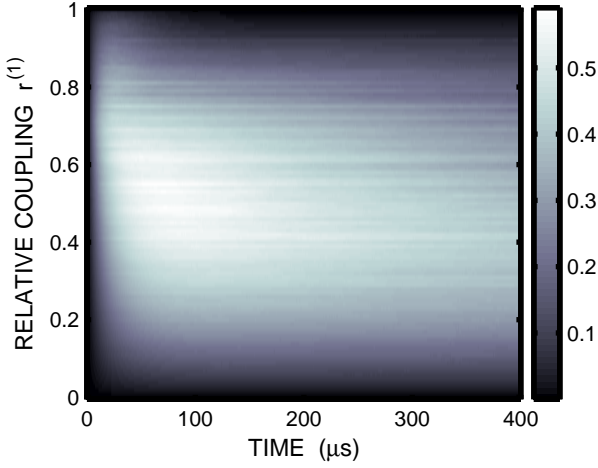}
\end{center}
\caption{\label{fig:weakCoupling} (Color online) Concurrence as a
function of time and the relative coupling strength $r^{(1)}$
in the weak ion-cavity coupling regime. The dynamics is initially
monotonic since the existing superradiant component decays rapidly
compared to other dynamical time scales. The subradiant state
component decays eventually because of the atomic spontaneous
emissions. The best entanglement production occurs with asymmetric couplings ($r^{(1)} \neq 1/\sqrt{2}$). Parameters: $\Delta / \Delta_0 = 100, \kappa / \kappa_0 = 0.1$}
\end{figure}

In this regime, the oscillatory dynamics stemming from the
coherent coupling between the atoms and the cavity is heavily
damped. In Fig.~\ref{fig:weakCoupling} we plot the concurrence as
a function of both time and the relative coupling strength $r^{(1)}$
for $\Delta / \Delta_0 = 100$ and $\kappa / \kappa_0 = 0.1$, giving $|g_\textrm{eff}|/ 2\pi = \xi g \Omega/2\pi\Delta = 17$~kHz. All the other parameters are chosen as in
Table~\ref{tab:physicalValues}. We recall that initially the atomic state $|\psi (0) \rangle = |1^{(1)} 0^{(2)}\rangle$ is factorized. The initial dynamics of the concurrence
shows a monotonic increase, as the superradiant component [see
Eq.~\eqref{eq:nonloso}] rapidly fades away while the subradiant
component remains intact. However, because of the presence of
spontaneous emission, the subradiant state is not anymore
perfectly decoupled from the dynamics and, consequently, the concurrence
will not reach a steady-state value.

We note that the best peak value of the concurrence $C\simeq0.6$
is achieved for $r^{(1)}\simeq0.55$, i.e., as expected, for an
asymmetric configuration ($r^{(1)} \neq 1/ \sqrt{2}$) of the ions with respect to the cavity
field. However, this value of  $r^{(1)}$ is now slightly different
than the one obtained in Sec.~\ref{sec:analytical} where
spontaneous emissions were neglected ($r^{(1)}=0.5$). We will
further discuss this point when considering the position
dependence of the jumps statistics at the end of this subsection.

\begin{figure}[tb]
\begin{center}
\includegraphics{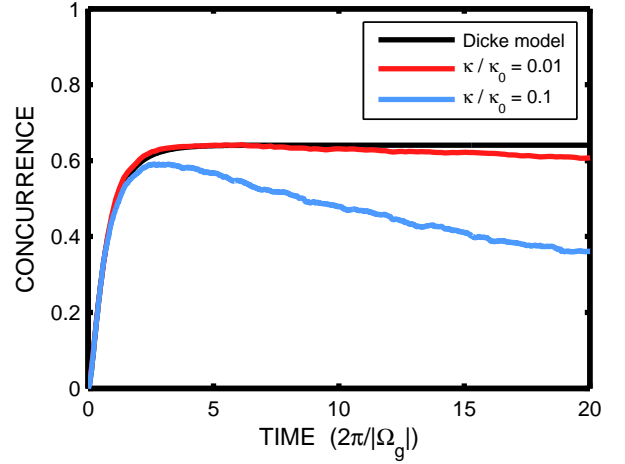}
\end{center}
\caption{\label{fig:weakCouplingSlides} (Color online) Dynamics of the concurrence in the weak ion-cavity coupling regime for $r^{(1)}=0.55$. In the Dicke model with cavity losses (highest line) a stationary value of the concurrence is reached as the superradiant component is over-damped. Parameters: $\Delta / \Delta_0 = 100, \kappa / \kappa_0 = 0.1$ (lowest line, $2\pi/|\Omega_g|=23$ $\mu$s); $\Delta / \Delta_0 = 1000, \kappa / \kappa_0 = 0.01$ (middle line, $2\pi/|\Omega_g|=230$ $\mu$s).}
\end{figure}

In Fig.~\ref{fig:weakCouplingSlides} we further study the effect
of the spontaneous emissions in the weak ion-cavity coupling case. In this
figure, we compare the predictions of the Dicke model, described
in Sec.~\ref{sec:nonIdealCavity}, with the dynamics of the
ion-cavity system in the presence of the spontaneous emissions for $\Delta / \Delta_0
= 100, \kappa / \kappa_0 = 0.1$ and  $\Delta / \Delta_0 = 1000, \kappa / \kappa_0 = 0.01$. The dynamics
of the concurrence clearly shows that, in the first case ($
\kappa / \kappa_0 = 0.1$), the system approximates the
Dicke model well while $|\Omega_g| t / 2\pi < 2.5 $, where
the generalized Rabi frequency $|\Omega_g|$ is given by Eq.~\eqref{eq:E}. For
a better cavity ($ \kappa / \kappa_0 = 0.01$), the concurrence
approaches its quasi-stationary value and the system approximates
the ideal Dicke dynamics for longer times, $|\Omega_g| t / 2\pi < 20$.

\begin{figure}[tb]
\begin{center}
\includegraphics{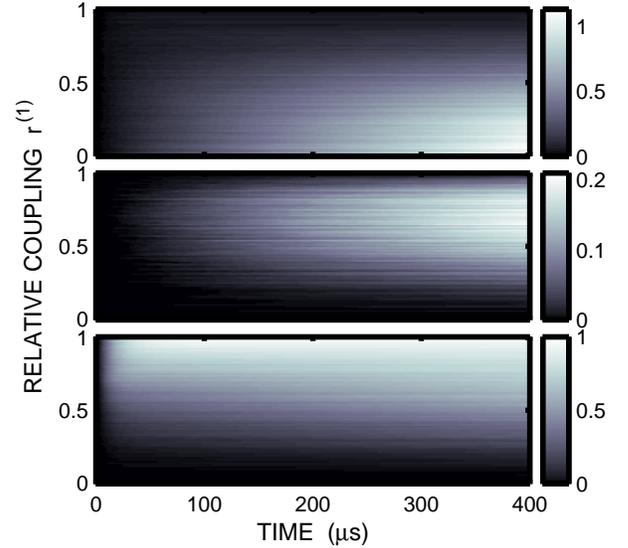}
\end{center}
\caption{\label{fig:weakCouplingJumps} (Color online) Average
cumulative number of quantum jumps per ensemble member for each
decay channel in the weak ion-cavity coupling regime. From above:
$C_S^{(1)}$, $C_S^{(2)}$, and $a$. Parameters: $\Delta / \Delta_0 = 100, \kappa / \kappa_0 = 0.1$.}
\end{figure}

We finally look at the statistics of the quantum jumps, described by the
jump operators $C_m^{(j)}$ in Eqs.~\eqref{eq:scaledC11}-\eqref{eq:scaledD12}.
First of all, we note that the source states $|\Phi_j\rangle$ [see Eqs.~\eqref{eq:decayingState1} and \eqref{eq:decayingState2}] of the jump operators $C_S^{(j)}$ and $C_D^{(j)}$ are identical for a given atom $j=1,2$. Therefore, the jump statistics of the two corresponding decay channels will
also be the same with a branching ratio given by  $\Gamma_S^{(j)} / \Gamma_D^{(j)} = \gamma_S / \gamma_D \simeq 13$. Our MCWF simulations confirm that the
dominant jump processes are those corresponding to the effective spontaneous emission operators
$C_S^{(j)}$ and the cavity photon annihilation operator $a$. In Fig.~\ref{fig:weakCouplingJumps} we plot
the average cumulative number of quantum jumps per ensemble member for the jump operators $C_S^{(1)}$, $C_S^{(2)}$, and $a$.

Looking at the statistics helps us to understand how the reservoir-mediated entanglement generation
process depends on $r^{(1)}$. We notice that the jump statistics of processes originating from the spontaneous
emissions of atoms 1 and 2 are different. This is of course due to the asymmetry in the initial condition. Since initially the excitation is present in
atom 1, the average cumulative number of jumps  per ensemble
member is typically greater for  $C_S^{(1)}$ than for $C_S^{(2)}$.
The peak in the cumulative number of jumps, for the three
different jump operators considered in
Fig.~\ref{fig:weakCouplingJumps}, moreover, is reached in
correspondence of different values of $r^{(1)}$. This indicates
that the value $r^{(1)}\simeq 0.55$, which optimizes the concurrence generation (see Fig.~\ref{fig:weakCoupling}), corresponds to a compromise
between the different  $r^{(1)}$-dependent jump statistics. In
particular, the deviation from the optimal value in the absence of spontaneous emission ($r^{(1)}=0.5$) might be due to the fact that the number of $C_S^{(1)}$-jumps increases for decreasing values of
$r^{(1)}$. 

\subsubsection{Strong ion-cavity coupling regime}

\begin{figure}[tb]
\begin{center}
\includegraphics{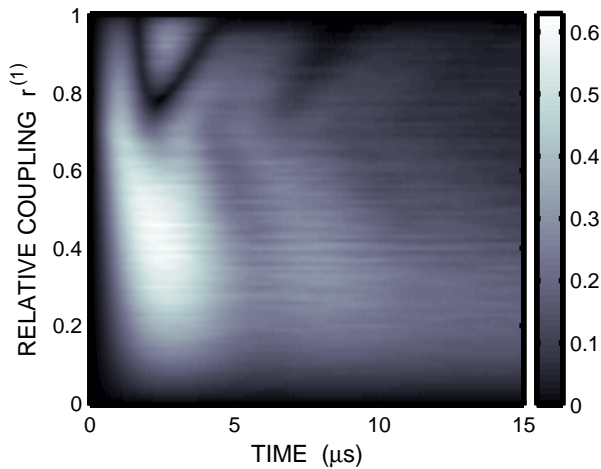}
\end{center}
\caption{\label{fig:strongCoupling} (Color online) Concurrence as
a function of time for different values of the relative coupling
strength $r^{(1)}$ in the strong ion-cavity coupling regime.
Oscillations appear because of a relative phase evolution between the
superradiant and subradiant states. Parameters: $\Delta / \Delta_0 = 10$ and $\kappa / \kappa_0 = 0.1$.}
\end{figure}

In the strong ion-cavity coupling regime, the cavity damping is slow
compared to the coherent dynamics. Therefore, a slowly damped
oscillatory behavior of the concurrence is expected. In
Fig.~\ref{fig:strongCoupling} we plot the concurrence as a
function of both time and the relative coupling strength $r^{(1)}$
for $\Delta / \Delta_0 = 10$ and $\kappa / \kappa_0 = 0.1$, giving $|g_\textrm{eff}|/ 2\pi = \xi g \Omega / 2\pi\Delta = 170$~kHz. All the other parameters are chosen as in
Table~\ref{tab:physicalValues}. Note that the ratio
$|g_\textrm{eff}|/\kappa$  is now one order of magnitude bigger
than in Sec.~\ref{sec:subsub}. The dynamics has an oscillatory
character, since the superradiant component survives much longer
than in the weak ion-cavity coupling regime. However, due to the presence of the spontaneous emissions the concurrence does not reach a steady-state value in this regime either.

\begin{figure}[tb]
\begin{center}
\includegraphics{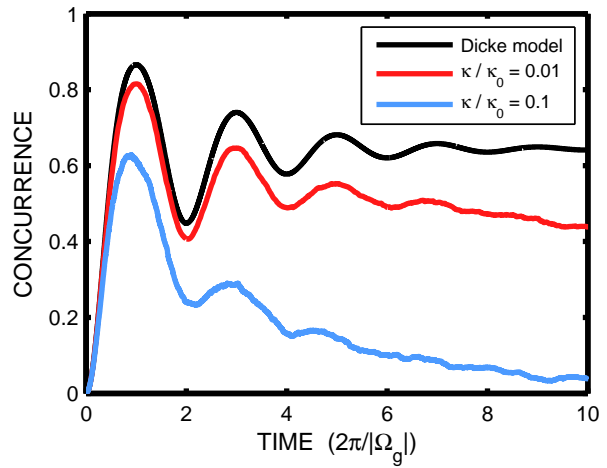}
\end{center}
\caption{\label{fig:strongCouplingSlides} (Color online) Time
evolution of the concurrence in the strong ion-cavity coupling regime
with relative coupling strength $r^{(1)}=0.46$. For the Dicke
model with cavity losses (highest line), the concurrence approaches
a constant stationary value after strong oscillations caused by
the slowly decaying superradiant component. Parameters: $\Delta / \Delta_0 =
10, \kappa / \kappa_0 = 0.1$ (lowest line, $2\pi/|\Omega_g|=2.7$ $\mu$s); $\Delta / \Delta_0 = 100, \kappa / \kappa_0 = 0.01$ (middle line, $2\pi/|\Omega_g|=27$ $\mu$s).}
\end{figure}

The best peak value of the concurrence, $C\simeq0.6$, is
now obtained for $r^{(1)}\simeq0.46$. In
Fig.~\ref{fig:strongCouplingSlides} we choose this value of
$r^{(1)}$ and we compare the dynamics of the single-mode Dicke
model with cavity losses to the dynamics of the ion-cavity system
in the presence of effective spontaneous emissions for the cases of
$\Delta / \Delta_0 = 10$ with $\kappa / \kappa_0 = 0.1$,
and $\Delta / \Delta_0 = 100$ with $\kappa / \kappa_0 = 0.01$.
In the second case, i.e., for a better quality factor, the system
approximates the Dicke model for longer time scales, as one
would expect. In this case one can clearly observe the damped Rabi
oscillation at the generalized Rabi frequency given by
Eq.~\eqref{eq:E}.

It is worth noticing that, in the strong ion-cavity coupling regime, the
laser-mediated interaction with the cavity vacuum allows us to
generate a highly entangled state of the two ions, as one can see
in  Fig.~\ref{fig:strongCouplingSlides}. In particular, for
$\Delta / \Delta_0 = 100$ with $\kappa / \kappa_0 = 0.01$,
using a laser pulse of duration $t \simeq 2\pi / |\Omega_g|$, the generated state is close to a maximally entangled Bell state.

\section{\label{sec:dispersiveRegime}
Environment-induced entanglement: Dispersive regime}

In the previous section we have seen that by placing the ions properly, i.e., by adjusting the relative coupling strength $r^{(1)}$, it is possible to optimize the reservoir-mediated entanglement generation. The
examples discussed above deal with the resonant effective model, which is defined by the condition $\delta_\textrm{eff} = 0$, which in turn corresponds to a physical detuning $\delta_L = \xi [ \Omega^2 - ( \beta_T g )^2 ]/ \Delta$ [cf.~Eq.~\eqref{eq:deltaEff}]. We have seen that the highest value of
the concurrence is obtained in the strong ion-cavity coupling regime.

In Ref.~\cite{Francica}, however, the single-mode Dicke model with
cavity losses is studied in the dispersive regime, showing that a
high degree of entanglement can be obtained also in the weak ion-cavity coupling regime. For this reason we now look at the off-resonant
entanglement generation process in the ion-cavity QED, i.e., we
consider the case in which $\delta_\textrm{eff} \neq 0$. In the
dispersive regime, the relative position of the ions does not play
an essential role and in fact one shows that the optimal value of
$r^{(1)}$  is obtained for equal coupling of the two ions, i.e.,
$r^{(1)} = r^{(2)} = 1/\sqrt{2}$~\cite{Francica}.

\begin{figure}[tb]
\begin{center}
\includegraphics{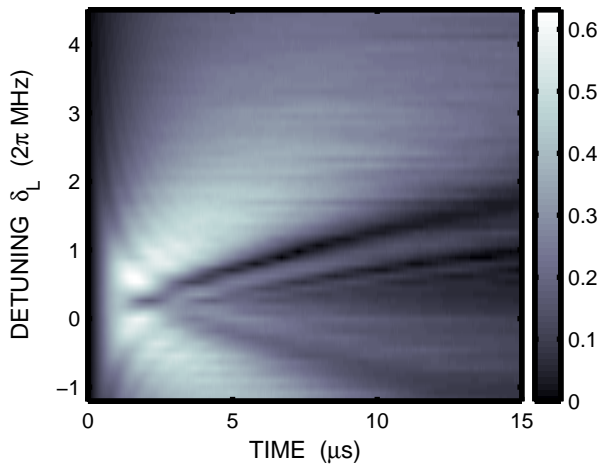}
\end{center}
\caption{\label{fig:dispersiveRegime} (Color online) Concurrence
as a function of time for different values of the detuning
$\delta_L$ for a homogeneously coupled case ($r^{(1)}=r^{(2)} = 1/\sqrt{2}$). The entanglement generation slows down when passing into the dispersive regime. The resonance is dislocated from the origin because of the Stark shifts. Parameters: $\Delta /\Delta_0 = 10$ and $\kappa / \kappa_0 = 0.1$. }
\end{figure}

We consider once more the initial atomic state $|\psi (0) \rangle =
|1^{(1)} 0^{(2)}\rangle$ combined with the cavity in the vacuum state
$|0^{(C)}\rangle$. We set $r^{(1)} = r^{(2)} = 1/\sqrt{2}$ (by choosing maximally strong cavity-driven couplings $\beta^{(1)} = \beta^{(2)} = 1$), $\Delta / \Delta_0 = 10$, and $\kappa / \kappa_0 = 0.1$,
corresponding to the weak ion-cavity coupling regime of
Sec.~\ref{sec:subsub}. We now look at the time evolution of the
concurrence for different values of the laser detuning $\delta_L$.
Figure~\ref{fig:dispersiveRegime} shows the concurrence as a
function of both time and detuning $\delta_L$. One can see clearly that the
Stark shift terms appearing in the effective Hamiltonian of
Eq.~\eqref{eq:Heffion} relocate the resonance condition from
the origin to  $\delta_L /2\pi = \xi [ \Omega^2 - ( \beta_T g )^2 ]/ 2\pi\Delta = 120$~kHz. Figure~\ref{fig:dispersiveRegime}
also shows that selecting the detuning $\delta_L$ further away from the resonance produces higher values of concurrence. In particular, with the chosen parameters the maximum value of concurrence $C \simeq 0.62$ is obtained with $\delta_L / 2\pi \simeq 600$~kHz.

As demonstrated in Ref.~\cite{Francica}, increasing the detuning $|\delta_\textrm{eff}|$ correspondingly increases the time it takes for the concurrence to
reach its peak value. The longer is the entanglement generation
time, however, the stronger is the effect of the spontaneous
emissions. In other words, the achieved gain in the entanglement
generation obtained by increasing the effective detuning is quickly suppressed due to the
spontaneous decay, as the overall time of the entanglement
generation process increases. The maximum value of entanglement
achievable in the dispersive regime is therefore determined by the
interplay between these two effects.

\begin{figure}[tb]
\begin{center}
\includegraphics{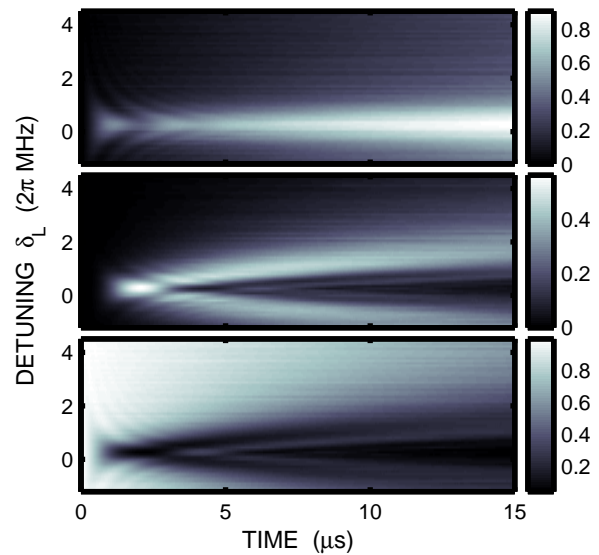}
\end{center}
\caption{\label{fig:dispersiveRegimePopulations} (Color online)
Populations of the atomic states $\rho_{00,00}$, $\rho_{01,01}$,
and $\rho_{10,10}$ (from above) as a function of time for
different values of detuning $\delta_L$. Parameters are as in Fig.~\ref{fig:dispersiveRegime}. }
\end{figure}

\begin{figure}[tb]
\begin{center}
\includegraphics{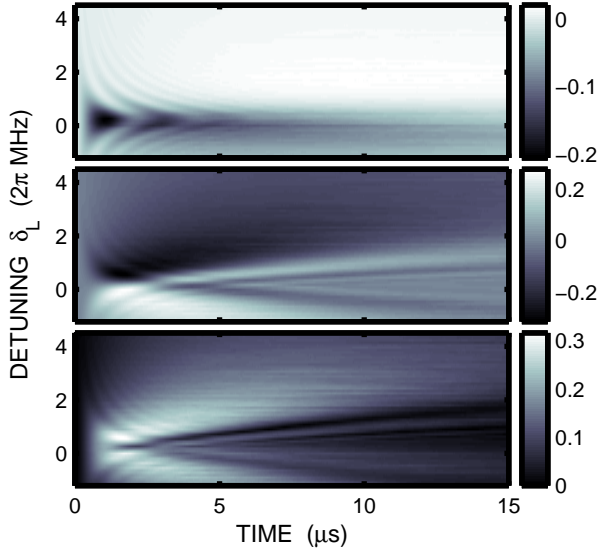}
\end{center}
\caption{\label{fig:dispersiveRegimeCoherences} (Color online)
Dynamics of $\textrm{Re}[\rho_{01,10}]$,
$\textrm{Im}[\rho_{01,10}]$, and $|\rho_{01,10}|$ (from above) as
a function of time for different values of detuning $\delta_L$
(observed concurrence is given by $C=2|\rho_{01,10}|$). Parameters are as in Fig.~\ref{fig:dispersiveRegime}.}
\end{figure}

It is worth noticing that going from the resonant into the
dispersive regime changes the character of the generated
entangled state as well. To illustrate this point, we plot in
Figs.~\ref{fig:dispersiveRegimePopulations}
and~\ref{fig:dispersiveRegimeCoherences}  the populations and
coherences, respectively, of the reduced atomic density matrix versus
time and detuning $\delta_L$. These plots confirm the increase in the
entanglement generation time when going deeper and deeper into the dispersive regime ($|\delta_\textrm{eff}| > 0$). If we then focus on the dynamics of the coherences and, in particular, on the real and imaginary parts of the only nonzero off-diagonal element $\rho_{01,10}$, we see that on resonance the imaginary part vanishes in accordance with the predictions of Sec.~\ref{sec:resonantRegime}. Therefore, in the resonant regime the generated entangled state approximates the subradiant state. On the other hand, in the dispersive regime Re$[\rho_{01,10}] \simeq 0$ and Im$[\rho_{01,10}]\neq 0$. Indeed, in the absence of the spontaneous emissions, the generated state in the dispersive regime would be $\left( \big| 1^{(1)} 0^{(2)} \big\rangle  \pm i \big| 0^{(1)} 1^{(2)} \big\rangle
\right)/\sqrt{2}$ (positive sign for negative $\delta_\textrm{eff}$ and vice versa).

\section{\label{sec:summary}
Summary and Conclusions}

In this paper we have investigated how the single-mode Dicke model
can be realized under experimentally feasible conditions using two
trapped $^{40}$Ca$^+$ ions inside a high-finesse optical cavity.
We have taken into account the spontaneous emissions of the ions
as well as the damping of the electromagnetic field inside the
cavity. In particular, we have derived an effective two-level
description of the three-level ions interacting with the cavity
mode.

We have shown that under suitable conditions the two ions indeed behave
collectively, with a coherent dynamical evolution well
described by the Dicke model: two effective two-level systems exchanging an excitation with an effective one-dimensional cavity mode. The presence of decohering processes, such as the atomic spontaneous emission or the cavity field damping, modifies this ideal picture. However, in the effective
model, the spontaneous emission decay rates are proportional to $1/\Delta^2$ whereas the ion-cavity couplings scale as $1 / \Delta$, where
$\Delta$ is the detuning of the physical cavity frequency from the electronic
transition that it is driving. This difference in the scaling can
be exploited in order to partly suppress the destructive effect of
the atomic spontaneous emissions.

We have identified the generation of entanglement as a fingerprint
of the cooperative atomic behavior and analyzed this process in
detail. In particular, we have proven that it is possible to enhance the entanglement generation process by positioning the ions appropriately at different locations with respect to the standing mode of the electromagnetic field inside the cavity. In the
resonant case, where the two-level systems and the cavity mode
have the same frequency, we have shown that asymmetric coupling
with the cavity mode produces the highest degree of entanglement,
even in presence of spontaneous emissions. We have studied both the
weak and the strong ion-cavity coupling regimes, defined by the strength of 
the ion-cavity excitation exchange compared to the cavity field damping rate, and found out the optimal conditions for entanglement generation in both cases.

Another possibility to optimize the entanglement generation is to
go to the dispersive regime in the ion-cavity coupling by using an off-resonant Raman
transition. The maximum degree of entanglement in the dispersive
and in the resonant regimes, for realistic values of the
parameters, is similar. Our results indicate, however, that the
character of the generated entangled state in the dispersive
regime changes compared to the resonant case.

Our experimental proposal is based on existing technology used in
the context of ion-cavity QED experiments
\cite{Walther,Piety,Matthias}.  In order to detect the generated
entanglement, the state tomography of the atomic systems is needed. In
recent years, this has been routinely performed in similar trapped-ion systems, e.g., in the context of quantum computation and measuring the quality of quantum gates \cite{Blatt09}. Therefore,
we expect our proposal to be within the reach of the experimental
community.

\acknowledgments

The authors thank K.-A. Suominen for useful discussions. S.M. thanks B. Garraway, M. Keller, and W. Lange for discussions on the experimental implementation of the ion-cavity QED setup and for the kind hospitality at the University of Sussex. This work was supported by the National Graduate School of Modern Optics and Photonics and the Magnus Ehrnrooth Foundation (K.H.), the Academy of Finland (Projects
No.~108699, No.~115682, No.~115982, and No.~8125004), the V\"ais\"al\"a Foundation, and the Turku Collegium of Science and Medicine (S.M.).

\appendix

\section{}
\label{app:adiabaticElimination}

In this appendix we show how the master
equation~\eqref{eq:model2me} for $\Lambda$-coupled three-level
atoms and a cavity photon is transformed into an effective
two-level master equation~\eqref{eq:meeffion} by adiabatic
elimination of the excited atomic states $\{ |P^{(j)} \rangle \}$.
Especially, the elimination transforms the jump operators related
to the spontaneous emissions into the form given
by Eqs.~\eqref{eq:jumpC1} and \eqref{eq:jumpC2}. For the sake of
generality, our treatment here is valid for $N$ atoms and we allow
each ion $j$ to be irradiated by a separate pump laser with frequency
$\omega_L + \delta_L^{(j)}$ and coupling strength $g_L^{(j)}$.

Passing into interaction picture $\varrho \mapsto \tilde{ \varrho} = e^{i K t } \varrho e^{-i K t}$ with respect to
\begin{equation}
K = \omega_C \Big( a^{\dag}a + \frac{1}{2} \Big) +  \sum_{j} \sum_{l=S,P,D}\omega_l A_{ll}^{(j)} - \Delta \sum_{j} A_{PP}^{(j)}
\end{equation}
transforms the operators as
\begin{align}
a & \mapsto \tilde{a} = e^{-i\omega_C t} a, \\
A_{PS}^{(j)} & \mapsto \tilde{A}_{PS}^{(j)} = e^{+i\omega_L t} A_{PS}^{(j)}, \\
A_{PD}^{(j)} & \mapsto \tilde{A}_{PD}^{(j)} = e^{+i\omega_C t} A_{PD}^{(j)}, \\
A_{DS}^{(j)} & \mapsto \tilde{A}_{DS}^{(j)} = e^{+i(\omega_L + \omega_C) t} A_{DS}^{(j)}, \\
A_{ll}^{(j)} & \mapsto \tilde{A}_{ll}^{(j)} = A_{ll}^{(j)}, \qquad l=S,P,D,
\end{align}
for all atoms $j$, and the Hamiltonian~\eqref{eq:model3H} becomes
\begin{equation}
\tilde{H} = \Delta \sum_j A_{PP}^{(j)} + \sum_j \Big( g_L^{(j)} e^{-i \delta_L^{(j)} t } A_{PS}^{(j)} + g_C^{(j)} a A_{PD}^{(j)} + H.c. \Big).
\end{equation}
The dissipator part of the master equation~\eqref{eq:model2me} is
invariant under this transformation.

Let us define a projection $\mathcal{P}$ to a subspace spanned by
the to-be-eliminated atomic states $\{ |P^{(j)} \rangle \}$, and another
projection $\mathcal{Q}$ to the complementary subspace by
\begin{align}
\mathcal{P} &\equiv \sum_j A_{PP}^{(j)} \otimes \mathbf{1}_\textrm{cav}, \\
\mathcal{Q} &\equiv \mathbf{1}  - \mathcal{P} = \sum_j \Big( A_{SS}^{(j)} + A_{DD}^{(j)} \Big) \otimes \mathbf{1}_\textrm{cav}.
\end{align}
Correspondingly, the density matrix divides into four sections
\begin{align}
\tilde{ \varrho } &=  \mathcal{Q} \tilde{ \varrho } \mathcal{Q} + \mathcal{Q} \tilde{ \varrho } \mathcal{P} + \mathcal{P} \tilde{ \varrho } \mathcal{Q} + \mathcal{P} \tilde{ \varrho } \mathcal{P} \nonumber \\
& \equiv \varrho_{QQ} + \varrho_{QP} + \varrho_{PQ} + \varrho_{PP}. \label{eq:rhoSections}
\end{align}
The Hamiltonian is similarly divided in parts
\begin{align}
H_{PP} &\equiv \mathcal{P} \tilde{H} \mathcal{P} = \Delta \sum_j A_{PP}^{(j)}, \label{eq:HPP}\\
H_{PQ} & = H_{QP}^\dagger \equiv \mathcal{P} \tilde{H} \mathcal{Q} \nonumber \\
& = \sum_j \Big( g_L^{(j)} e^{-i\delta_L^{(j)} t} A_{PS}^{(j)} + g_C^{(j)} a A_{PD}^{(j)} \Big), \label{eq:HPQ} \\
H_{QQ} &\equiv \mathcal{Q} \tilde{H} \mathcal{Q} = 0. \label{eq:HQQ}
\end{align}

We proceed to deriving an effective master equation for
$\varrho_{QQ}$, which describes the dynamics of a collection of
effective two-level atoms and a cavity mode. Applying
Eqs.~\eqref{eq:rhoSections}-\eqref{eq:HQQ} to master
equation~\eqref{eq:model2me} gives
\begin{align}
\dot{ \varrho }_{QQ} = & -i H_{QP} \varrho_{PQ} + i \varrho_{QP} H_{PQ} -\frac{\kappa }{2} \Big( \{ a^\dagger a, \varrho_{QQ} \}  \nonumber \\
& - 2 a \varrho_{QQ} a^\dagger \Big) + \gamma_S \sum_j A_{SP}^{(j)} \varrho_{PP} A_{PS}^{(j)} \nonumber \\ 
& + \gamma_D \sum_j A_{DP}^{(j)} \varrho_{PP} A_{PD}^{(j)}, \label{eq:rhoQQ} \\
\dot{ \varrho }_{QP} = &\,\,  \dot{ \varrho }_{PQ}^\dagger = -\Big( \frac{\gamma_S + \gamma_D }{2} - i \Delta \Big) \varrho_{QP} + i \varrho_{QQ} H_{QP} \nonumber \\ 
&  - i H_{QP} \varrho_{PP} - \frac{\kappa }{2} \Big( \{ a^\dagger a, \varrho_{QP} \} - 2 a \varrho_{QP} a^\dagger \Big), \label{eq:rhoQP}\\
\dot{ \varrho }_{PP} = & -(\gamma_S + \gamma_D ) \varrho_{PP} - i H_{PQ} \varrho_{QP} + i \varrho_{PQ} H_{QP} \nonumber \\ 
&-\frac{\kappa }{2} \Big( \{ a^\dagger a, \varrho_{PP} \} - 2 a \varrho_{PP} a^\dagger \Big). \label{eq:rhoPP}
\end{align}
Setting $\dot{ \varrho }_{QP} = \dot{ \varrho }_{PQ} = 0$, assuming $\varrho_{QQ} \gg \varrho_{PP}$, and neglecting the cavity damping in Eq.~\eqref{eq:rhoQP} gives an approximation
\begin{align}
\varrho_{QP} &= \varrho_{PQ}^\dagger \simeq \frac{ -\Delta + i(\gamma_S + \gamma_D )/2 }{\Delta^2 + (\gamma_S +\gamma_D )^2 /4 } \varrho_{QQ} H_{QP}. \label{eq:rhoQP_approx}
\end{align}
Similarly, setting $\dot{ \varrho }_{PP} = 0$ in
Eq.~\eqref{eq:rhoPP} and using the above approximations for
$\varrho_{QP}$ and $\varrho_{PQ}$ gives
\begin{align}
\varrho_{PP} &\simeq -\frac{ i }{ \gamma_S + \gamma_D } ( H_{PQ} \varrho_{QP} - \varrho_{PQ} H_{QP} )  \nonumber \\
& \simeq \frac{1}{\Delta^2 + (\gamma_S + \gamma_D )^2/4} H_{PQ} \varrho_{QQ} H_{QP}.\label{eq:rhoPP_approx}
\end{align}
Finally, by inserting Eqs.~\eqref{eq:rhoQP_approx} and \eqref{eq:rhoPP_approx} into Eq.~\eqref{eq:rhoQQ}, we arrive at an approximated master equation
\begin{widetext} 
\begin{align}
\dot{ \varrho }_{QQ} \simeq & -i \Big[ \frac{-\Delta }{\Delta^2 + (\gamma_S + \gamma_D )^2 /4} H_{QP} H_{PQ}, \varrho_{QQ} \Big] -\frac{\kappa }{2} \Big( \{ a^\dagger a, \varrho_{QQ} \} - 2 a \varrho_{QQ} a^\dagger \Big) \nonumber \\
& - \sum_j \frac{1}{2} \frac{\gamma_S}{\Delta^2 + (\gamma_S + \gamma_D )^2 /4} \Big[ \big\{ ( H_{QP} A_{PS}^{(j)} ) ( A_{SP}^{(j)} H_{PQ} ), \varrho_{QQ} \big\} -2 (A_{SP}^{(j)} H_{PQ}) \varrho_{QQ} (H_{QP} A_{PS}^{(j)}) \Big] \nonumber \\
& - \sum_j \frac{1}{2} \frac{\gamma_D}{\Delta^2 + (\gamma_S + \gamma_D )^2 /4} \Big[ \big\{ ( H_{QP} A_{PD}^{(j)} ) ( A_{DP}^{(j)} H_{PQ} ), \varrho_{QQ} \big\} -2 (A_{DP}^{(j)} H_{PQ}) \varrho_{QQ} (H_{QP} A_{PD}^{(j)}) \Big] \nonumber \\
\equiv & -i \left[ H_{\rm eff}, \varrho_{QQ}\right] - \frac{\kappa}{2} \left( \{ a^{\dag}a, \varrho_{QQ} \} -2a \varrho_{QQ} a^{\dag} \right) - \sum_j \sum_{m=S,D} \frac{\Gamma_m^{(j)}}{2} \Big[ \big\{ C_m^{(j)\dag}C_m^{(j)}, \varrho_{QQ} \big\} -2  C_m^{(j)} \varrho_{QQ}  C_m^{(j)\dag} \Big],
\end{align}
\end{widetext}
which has the form of the master equation~\eqref{eq:meeffion}. We can now recognize the effective Hamiltonian [cf.~Eq.~\eqref{eq:Heffion}] as
\begin{align}
H_\textrm{eff} \equiv & - \frac{\Delta }{\Delta^2 + (\frac{\gamma_S + \gamma_D}{2})^2} H_{QP} H_{PQ} \nonumber \\
= & - \xi \sum_j \Big[ \Big( e^{-i \delta_L^{(j)} t} \frac{g_C^{(j)*} g_L^{(j)} }{\Delta } \, a^\dagger A_{DS}^{(j)} + h.c. \Big) \nonumber \\
& + \frac{|g_L^{(j)}|^2}{\Delta } A_{SS}^{(j)} + \frac{|g_C^{(j)}|^2 }{\Delta } a^\dagger a A_{DD}^{(j)} \Big],
\end{align}
the effective spontaneous emission jump operators [cf.~Eqs.~\eqref{eq:jumpC1} and \eqref{eq:jumpC2}] as
\begin{align}
C_S^{(j)} &= A_{SP}^{(j)} H_{PQ} = g_L^{(j)} e^{-i\delta_L^{(j)} t } A_{SS}^{(j)} + g_C^{(j)} a A_{SD}^{(j)}, \\
C_D^{(j)} &= A_{DP}^{(j)} H_{PQ} = g_L^{(j)} e^{-i\delta_L^{(j)} t } A_{DS}^{(j)} + g_C^{(j)} a A_{DD}^{(j)},
\end{align}
(N.B., the operators are unique up to a global phase factor
$e^{i\theta_m^{(j)}}$, where $\theta_m^{(j)} \in \mathbb{R}$), and the
corresponding decay rates as
\begin{align}
\Gamma_S^{(j)} &= \frac{ \gamma_S }{ \Delta^2 + (\gamma_S + \gamma_D )^2 /4 } = \xi \, \frac{\gamma_S }{\Delta^2 }, \\
\Gamma_D^{(j)} &= \frac{ \gamma_D }{ \Delta^2 + (\gamma_S + \gamma_D )^2 /4 } = \xi \, \frac{\gamma_D }{\Delta^2 }.
\end{align}
In the above equations the dimensionless prefactor $\xi$ is as defined in Eq.~\eqref{eq:xi}.

\section{}
\label{app:phaseRotation}

In this appendix, we review how the effective two-level Hamiltonian~\eqref{eq:Heffion} with Stark shifts is matched exactly with the Tavis-Cummings Hamiltonian~\eqref{eq:HDicke2} by passing into a rotating frame. Moreover, we show how to exploit the same phase transformation in order to simplify
the propagator for the numerical simulations. As in Appendix~\ref{app:adiabaticElimination}, we assume each atom
$j=1, \ldots , N$ to be addressed by its own pump laser with
independent detunings $\delta_L^{(j)}$ and couplings $g_L^{(j)}$.

We will restrict ourselves to the subspace with at most one
excitation. We introduce the following notation for the atomic
states:
\begin{align}
|\varphi_0 \rangle &= |0^{(1)} \cdots 0^{(N)} \rangle, \\
|\varphi_1^{(j)} \rangle &= |0^{(1)} \cdots 1^{(j)} \cdots 0^{(N)}
\rangle ,
\end{align}
so that the Hamiltonian~\eqref{eq:Heffion} reads
\begin{align}
H = & \sum_j \Big[ \big( e^{-i \delta_L^{(j)} t } \lambda^{(j)} |\varphi_0 \, 1^{(C)}\rangle \langle \varphi_1^{(j)}\, 0^{(C)}| + H.c. \big) \nonumber \\
& + S^{(j)} |\varphi_1^{(j)} \, 0^{(C)}\rangle \langle \varphi_1^{(j)} \, 0^{(C)}| \Big] \nonumber \\
& + S^{(C)} |\varphi_0 \, 1^{(C)}\rangle \langle \varphi_0 \, 1^{(C)}|,
\end{align}
with Stark shifts $S^{(C)} = -\xi \sum_j |g_C^{(j)}|^2 /\Delta$ and $S^{(j)} =
-\xi |g_L^{(j)}|^2 / \Delta$, and effective ion--cavity couplings $\lambda^{(j)} = -\xi g_C^{(j)*}
g_L^{(j)} / \Delta$. The time dependence of the coupling terms
will be eliminated by a phase rotation of the basis vectors
\begin{align}
|\varphi_0 \, 0^{(C)} & \rangle \mapsto e^{i\mu t} |\varphi_0 \, 0^{(C)} \rangle,\\
|\varphi_0 \, 1^{(C)} & \rangle \mapsto e^{i\nu t} |\varphi_0 \, 1^{(C)} \rangle,\\
|\varphi_1^{(j)} \, 0^{(C)} & \rangle \mapsto e^{i(\delta_L^{(j)} + \nu )t } |\varphi_1^{(j)} \, 0^{(C)} \rangle,
\end{align}
where $\mu, \nu \in \mathbb{R}$ are up to now free parameters. The
Hamiltonian transforms accordingly into
\begin{align}
H \mapsto H' = & \sum_j \Big[ \big( \lambda^{(j)} |\varphi_0 \, 1^{(C)}\rangle \langle \varphi_1^{(j)} \, 0^{(C)}| + H.c. \big) \nonumber \\
& + ( S^{(j)} + \delta_L^{(j)} + \nu ) |\varphi_1^{(j)} \, 0^{(C)}\rangle \langle \varphi_1^{(j)} \, 0^{(C)}| \Big] \nonumber \\
& + ( S^{(C)} + \nu )|\varphi_0 \, 1^{(C)}\rangle \langle \varphi_0 \, 1^{(C)}| \nonumber \\
& + \mu |\varphi_0 \, 0^{(C)}\rangle \langle \varphi_0 \, 0^{(C)}|
\end{align}

The requirement of a full compatibility with the Tavis-Cummings Hamiltonian~\eqref{eq:HDicke2} within our restricted Hilbert space demands
that $\mu = \mu( \nu ) = ( S^{(C)} + \nu ) / 3$. Consequently, the
effective Dicke model parameters are identified as [cf.
Eqs.~\eqref{eq:omegaCEff}-\eqref{eq:alphaEff}]
\begin{align}
\omega_C^\textrm{eff} & = \frac{2}{3} S^{(C)} + \frac{2}{3} \nu , \\
\omega_A^{\textrm{eff} (j)} & = \delta_L^{(j)} + S^{(j)} - \frac{1}{3} S^{(C)} + \frac{2}{3} \nu, \\
\alpha_\textrm{eff}^{(j)} & = \lambda^{(j)}.
\end{align}
Moreover, the detunings are [cf.~Eq.~\eqref{eq:deltaEff}]
\begin{equation}
\delta_\textrm{eff}^{(j)} = \omega_A^{\textrm{eff}(j)} - \omega_C^\textrm{eff} = \delta_L^{(j)} + S^{(j)} - S^{(C)}.
\end{equation}
This means that the resonance condition of the Dicke model (cf.~Sec.~\ref{sec:resonantRegime}) is achieved with laser
detunings $\delta_L^{(j)}  = S^{(C)} - S^{(j)}$, while in the dispersive regime (cf.~Sec.~\ref{sec:dispersiveRegime}) $\delta_L^{(j)} \neq S^{(C)} - S^{(j)}$. As a conclusion,
having each (identical) atom driven by their own pump laser allows us
to simulate the inhomogeneous Dicke model~\eqref{eq:HDicke2} with
independent two-level transition frequencies $\omega_A^{(j)}$ and couplings
$\alpha^{(j)}$. On the other hand, if one has only a single laser
driving all of the atoms, the transition frequencies are the same for every atom 
$j$, but the coupling constants $\alpha^{(j)}$ remain independent
because of the position-dependent cavity couplings $g_C^{(j)}$.

Another aspect of the performed phase transformation is provided by how they affect the dissipator part of the master
equation~\eqref{eq:meeffion}. Within our restricted Hilbert space, the jump operators transform now as
\begin{align}
C_S^{(j)} &\mapsto e^{-i(\delta_L^{(j)} + \nu ) t } |\varphi_1^{(j)} \, 0^{(C)} \rangle \langle \Phi_j | , \\
C_D^{(j)} &\mapsto e^{-i(\delta_L^{(j)} + \nu ) t } |\varphi_0 \, 0^{(C)} \rangle \langle \Phi_j | , \\
a &\mapsto e^{-i [\mu(\nu) - \nu ] t} |\varphi_0 \, 0^{(C)}\rangle \langle \varphi_0 \, 1^{(C)}|
\end{align}
(global phase factors can be discarded immediately), where the
decaying (un-normalized) states are
\begin{align}
|\Phi_j \rangle = e^{ -i \nu t } g_L^{(j)* } \, |\varphi_1^{(j)} \, 0^{(C)}\rangle + g_C^{(j)*} |\varphi_0 \, 1^{(C)}\rangle .
\end{align}

In the numerical MCWF simulations, the dynamics is generated by a
non-Hermitian Monte Carlo Hamiltonian $H_\textrm{MC} = H -
\frac{i}{2} \sum_m \Delta_m J_m^\dagger J_m$, where $H$ is the
Hermitian Hamiltonian of the master equation, and $J_m$ and
$\Delta_m$ are all the jump operators and corresponding decay rates picked up
from the dissipator part of the master equation. From the
practical point of view, it is advantageous to have a
time-independent $H_\textrm{MC}$, since then the Dyson series of
the propagator simplifies into exponential form $U(t,t_0) = \exp [
- i H_\textrm{MC} (t - t_0)]$. This is now achieved simply by
choosing $\nu = 0$, and hence the phase transformation is unique.


\begin{thebibliography}{99}

\bibitem{Dicke}
R. H. Dicke, Phys. Rev. \textbf{93}, 99 (1954).

\bibitem{HarochePR}
M. Gross and S. Haroche, Phys. Rep. \textbf{93}, 301 (1982).

\bibitem{tanas02}
Z. Ficek and R. Tanas,  Phys. Rep.  \textbf{372}, 369 (2002).

\bibitem{IonsRev}
D. Leibfried, R. Blatt, C. Monroe, and D. Wineland, Rev. Mod. Phys. \textbf{75}, 281 (2003).

\bibitem{Bloch08}
I. Bloch, Nature (London) \textbf{453}, 1016 (2008).

\bibitem{DeVoe}
R.G. De Voe and R.G. Brewer, Phys. Rev. Lett. \textbf{76}, 2049 (1996).

\bibitem{HarocheSuperR}
J.M. Raimond, P. Goy, M. Gross, C. Fabre, and S. Haroche, Phys. Rev. Lett. \textbf{49}, 1924 (1982).

\bibitem{Walther}
G. R. Guth\"{o}hrlein, M. Keller, K. Hayasaka, W. Lange, and H. Walther, Nature (London) \textbf{414}, 49 (2001); 
M. Keller, B. Lange, K. Hayasaka, W. Lange, and H. Walther, Nature (London) \textbf{431}, 1075 (2004).

\bibitem{Piety}
A. B. Mundt, A. Kreuter, C. Becher, D. Leibfried, J. Eschner, F. Schmidt-Kaler, and R. Blatt, Phys. Rev. Lett. \textbf{89}, 103001 (2002); 
A. Kreuter, C. Becher, G. P. T. Lancaster, A. B. Mundt, C. Russo, H. H\"{a}ffner, C. Roos, J. Eschner, F. Schmidt-Kaler, and R. Blatt, Phys. Rev. Lett. \textbf{92}, 203002 (2004).

\bibitem{Benivegna}
G. Benivegna and A. Messina, J. Mod. Opt.  \textbf{36}, 1205 (1989).

\bibitem{BenivegnaPL}
G. Benivegna and A. Messina, Phys. Lett. A \textbf{126}, 249 (1988).

\bibitem{BuzekZei}
V. Bu\v{z}ek, Z. Phys. D \textbf{17}, 91 (1990).

\bibitem{Herskind2009}
P. F. Herskind, A. Dantan, J. P Marler, M. Albert, and M. Drewsen, Nature Phys. \textbf{5}, 494 (2009).

\bibitem{TCmodel}
M. Tavis and F. W. Cummings, Phys. Rev. \textbf{170}, 379 (1968).

\bibitem{Kimble}
H. J. Kimble, Nature (London) \textbf{453}, 1023 (2008).

\bibitem{Pellizzari}
T. Pellizzari, S. A. Gardiner, J. I. Cirac, and P. Zoller, Phys. Rev. Lett. \textbf{75}, 3788 (1995).

\bibitem{vanEnk}
S. J. van Enk, J. I. Cirac, and P. Zoller, Phys. Rev. Lett. \textbf{79}, 5178 (1997).

\bibitem{Plenio}
M. B. Plenio, S. F. Huelga, A. Beige, and P. L. Knight, Phys. Rev. A \textbf{59}, 2468 (1999).

\bibitem{Zheng}
S.-B. Zheng and G.-C. Guo, Phys. Rev. Lett. \textbf{85}, 2392 (2000).

\bibitem{Pachos}
J. Pachos and  H. Walther, Phys. Rev. Lett. \textbf{89}, 187903 (2002).

\bibitem{Lougovski}
P. Lougovski, E. Solano, and H. Walther, Phys. Rev. A \textbf{71}, 013811 (2005).

\bibitem{Chimczak}
G. Chimczak, R. Tana\'{s}, and A. Miranowicz, Phys. Rev. A \textbf{71}, 032316 (2005).

\bibitem{Li}
S.-B. Li and J.-B. Xu, Phys. Rev. A \textbf{72}, 022332 (2005).

\bibitem{Li07}
S.-B. Li, Phys. Rev. A \textbf{75}, 054304 (2007).

\bibitem{Chimczak08}
G. Chimczak and R. Tana\'{s}, Phys. Rev. A \textbf{77}, 032312 (2008).

\bibitem{Bina}
M. Bina, F. Casagrande, A. Lulli, and E. Solano, Phys. Rev. A \textbf{77}, 033839 (2008).

\bibitem{Natali2007}
S. Natali and Z. Ficek, Phys. Rev. A \textbf{75}, 042307 (2007).

\bibitem{ESDLaura}
L. Mazzola, S. Maniscalco, J. Piilo, K.-A. Suominen, and B. M. Garraway, Phys. Rev. A \textbf{79}, 042302 (2009).

\bibitem{Man08}
S. Maniscalco, F. Francica, R. L. Zaffino, N. Lo Gullo, and F. Plastina, Phys. Rev. Lett. \textbf{100}, 090503 (2008).

\bibitem{Matthias}
M. Keller, B. Lange, K. Hayasaka, W. Lange, and H. Walther, New J. Phys. \textbf{6}, 95 (2004).

\bibitem{Blatt09}
T. Monz, K. Kim, W. H\"ansel, M. Riebe, A. S. Villar, P. Schindler, M. Chwalla, M. Hennrich, and R. Blatt,
Phys. Rev. Lett. \textbf{102}, 040501 (2009).

\bibitem{DiFidio}
C. Di Fidio, S. Maniscalco, W. Vogel and A. Messina, Phys. Rev. A \textbf{65}, 033825 (2002).

\bibitem{MCWF} 
J. Dalibard, Y. Castin, and K. M{\o}lmer, Phys. Rev. Lett. \textbf{68}, 580 (1992).

\bibitem{ZollerCarmichael}
R. Dum, P. Zoller, and H. Ritsch, Phys. Rev. A \textbf{45}, 4879 (1992);
R. Dum, A. S. Parkins, P. Zoller, and C. W. Gardiner, Phys. Rev. A \textbf{46}, 4382 (1992);
H. Carmichael, \textsl{An Open System Approach to Quantum Optics}, Lecture Notes in Physics (Springer-Verlag, Berlin, 1993), Vol. m18.

\bibitem{wootte}
W. K. Wootters, Phys. Rev. Lett. \textbf{80}, 2245 (1998).

\bibitem{W}
W. Lange (private communication).

\bibitem{Francica}
F. Francica, S. Maniscalco, J. Piilo,  F. Plastina, and K.-A. Suominen, Phys. Rev. A \textbf{79}, 032310 (2009).

\end{thebibliography}
\end{document}